\documentclass[aps,reprint,floatfix]{revtex4-1}
\pdfoutput=1

\usepackage[T1]{fontenc}
\usepackage[utf8]{inputenc}
\usepackage{lmodern}

\usepackage[colorlinks=true,allcolors=blue]{hyperref}
\usepackage{xcolor}
\usepackage{ulem}

\usepackage{mathtools}
\usepackage{physics}

\newcommand*{\Rg}{R_\text{g}}

\newcommand*{\kt}{k_\text{B}T}
\newcommand*{\lp}{l_\text{p}}
\newcommand*{\fa}{f_\text{a}}

\renewcommand*{\vb}[1]{\boldsymbol{\mathrm{#1}}}
\renewcommand*{\vu}[1]{\hat{\vb{#1}}}

\begin{document}

\title{Statistical properties of a tangentially driven active filament}
\author{Matthew S. E. Peterson}
\author{Michael F. Hagan}
\author{Aparna Baskaran}
\affiliation{
    Martin A. Fisher School of Physics,
    Brandeis University, Waltham, MA, 02453
}
\date{\today}

\begin{abstract}
Active polymers play a central role in many biological systems, from bacterial
flagella to cellular cytoskeletons. Minimal models of semiflexible active
filaments have been used to study a variety of interesting phenomena in active
systems, such as defect dynamics in active nematics, clustering and laning in
motility assays, and conformational properties of chromatin in eukaryotic cells.
In this paper, we map a semiflexible polymer to an exactly solvable active Rouse
chain, which enables us to analytically compute configurational and dynamical
properties of active polymers with arbitrary rigidity. Upon mapping back to the
semiflexible filament, we see that the center of mass diffusion coefficient
grows linearly with an activity parameter that is renormalized by the polymer
persistence length. These results closely agree with numerical data obtained
from microscopic simulations.
\end{abstract}

\maketitle


\section{Introduction}

Active systems are characterized by constituents that consume energy to
produce directed motion. These systems are inherently out-of-equilibrium, and
thus lead to novel steady state behaviors~\cite{ramaswamy2010, marchetti2013}.
For example, simple model systems such as active nematics composed of
microtubules driven by kinesin motor proteins can continuously create and
annihilate topological defects~\cite{sanchez2012, giomi2013, decamp2015,
doostmohammadi2018, thampi2014}, and active Brownian particles can aggregate
into ``active solids''~\cite{redner2013, bechinger2016}.

Active polymers are a class of active systems that are of considerable interest
due to their prevalence in biological systems on multiple length scales,
including the flagella of bacterial microswimmers~\cite{chelakkot2014,
elgeti2015}, chromatin in eukaryotic cells~\cite{bronstein2009, bronshtein2015,
cabal2006, zidovska2013, ganai2014,haddad2017perspectives}, and actin in
cellular cytoskeletons~\cite{brangwynne2008, mizuno2007}. Many studies have
focused on the collective dynamics of many such filaments~\cite{le_goff2001,
prathyusha2018, weber2015, humphrey2002, sakaue2017, suzuki2017, sokolov2007,
saintillan2018}, and found that activity can lead to behaviors such as formation
of clusters~\cite{suzuki2017} and spiral patterns~\cite{gupta2019,
prathyusha2018, isele-holder2015}.

As in the context of self-propelled particle models \cite{ten2011brownian,
angelani2011active, ai2013rectification, fily2014dynamics, basu2018,
dauchot2019, kranz2019}, it is fruitful to understand the properties of isolated
active units to provide a framework for understanding the non-equilibrium steady
states that emerge in these complex systems. While there are few analytical
results available to date~\cite{liverpool2003, gao2017, eisenstecken2016,
eisenstecken2017, osmanovic2017, osmanovic2018, de_canio2017}, a number of
numerical studies have been undertaken to understand the statistical properties
of single active filaments~\cite{bronshtein2015, bronstein2009, weber2010,
ghosh2014, osmanovic2017, osmanovic2018, eisenstecken2016, di_pierro2018,
bianco2018, chaki2019, anand2018, bianco2018, das2019}. Filaments placed in a
bath of active particles can have anomalous dynamic properties, including super-
and sub-diffusive motion~\cite{bronshtein2015, bronstein2009, weber2010,
ghosh2014, osmanovic2017, osmanovic2018}, as well as enhanced diffusion
coefficients~\cite{eisenstecken2016, di_pierro2018, bianco2018, chaki2019}.
Activity was also found to lead to the ``softening'' of semiflexible filaments,
effectively reducing the persistence length, while sufficiently large active
forces could lead to chain swelling~\cite{eisenstecken2017, harder2014}. Similar
results have been found in the case where the active forces are directed along
the filament tangent~\cite{anand2018, bianco2018}, such as in actin or
microtubule motility assays.

In this work, we seek to understand the influence of activity on the statistical
properties of an isolated semiflexible filament subject to tangential active
forces~\cite{gupta2019, bianco2018, anand2018}, using an analytically tractable
model. Note that, neglecting excluded volume interactions, a semiflexible
filament with persistence length $\lp$ can be modeled as a Rouse chain with bond
length $b \approx 2 \lp$~\cite{doi2007}. Motivated by this mapping, we consider
a single active Rouse chain with activity directed along the tangent. We show
analytically that activity leads to an enhanced diffusion coefficient that grows
\textit{linearly} with the strength of the active force, while the end to end
distance of the polymer is \textit{independent} of activity. Mapping the typical
Rouse bond length, $b$, to the persistence length, $\lp$, we obtain an
analytical expression for the diffusion coefficient of an active semiflexible
polymer. We compare these predictions to Langevin dynamics simulations of both
Rouse chains and semiflexible filaments, and find that our analytical results
are able to accurately describe both cases. These results are directly relevant
for motility assay experiments~\cite{pringle2013, farhadi2018, sumino2012,
schaller2010, suzuki2017}, and elucidate behaviors of active units with internal
degrees of freedom.


\section{The active Rouse model}
A Rouse chain is a simple polymer model wherein we have $N$ beads connected by
harmonic bonds. Assuming this chain is in a highly viscous medium with thermal
noise, we obtain the familiar Rouse equation of motion for the $n$th
bead~\cite{doi2007}
\begin{equation} \label{eq:BasicRouseEOM}
    \gamma \pdv{\vb{r}_n}{t} = k (\vb{r}_{n+1} + \vb{r}_{n-1} - 2 \vb{r}_n)
    + \sqrt{2 \gamma \kt} \vb{\xi}_n(t),
\end{equation}
where $\gamma$ is the damping coefficient, and $k$ the spring constant. Note
that the Rouse chain simply collapses to a point in the zero temperature limit.
A non-zero temperature is necessary to give the polymer a finite size. The
root-mean-square (RMS) bond length is given by $b_0^2 = d \kt/k$, where $d$ is
the system dimensionality. The thermal noise $\vb{\xi}_n(t)$ is Gaussian white
noise with moments
\begin{eqnarray}
    \ev{\vb{\xi}_n(t)} &=& 0 \quad \text{and} \nonumber \\
    \ev{\xi_n^\alpha(t) \xi_m^\beta (t')}
        &=& \delta_{\alpha \beta} \delta_{nm} \delta(t-t').
\end{eqnarray}
Eq.~\eqref{eq:BasicRouseEOM} holds for $n = 2,\ldots,N-1$. At the ends,
we have
\begin{eqnarray*}
    \gamma \pdv{\vb{r}_1}{t} &=& k (\vb{r}_2 - \vb{r}_1)
    + \sqrt{2 \gamma \kt} \vb{\xi}_1(t), \\
    \gamma \pdv{\vb{r}_N}{t} &=& k (\vb{r}_{N-1} - \vb{r}_N)
    + \sqrt{2 \gamma \kt} \vb{\xi}_N(t).
\end{eqnarray*}
However, we can extend Eq.~\eqref{eq:BasicRouseEOM} to hold for all $n$ provided
we allow for ``ghost'' beads such that $\vb{r}_0 = \vb{r}_1$ and $\vb{r}_{N+1} =
\vb{r}_N$. This Rouse model describes an idealized filament in a dry system and
has served as an important model for obtaining physical intuition about the
statistical properties of polymers~\cite{doi2007}.

We add tangential activity to this polymer by supposing that the bonds of the
polymer impart a force on their attached beads. That is, if the $n$th bond
connects beads $n$ and $n+1$, then each of those beads experiences some force
$\vb{A}_n/2$ (so that the total force generated by the bond is $\vb{A}_n$). We
consider the simple case where
\[
    \vb{A}_n = \fa \times (\vb{r}_{n+1} - \vb{r}_n),
\]
with $f_a$ a constant parameter, in which the equation of motion for an active
Rouse chain is
\begin{eqnarray} \label{eq:DiscreteRouseEOM}
    \gamma \pdv{\vb{r}_n}{t} &=& k (\vb{r}_{n+1} + \vb{r}_{n-1} - 2 \vb{r}_n)
    + \fa \left(\frac{\vb{r}_{n+1} - \vb{r}_{n-1}}{2}\right)
    \nonumber \\
    && + \sqrt{2 \gamma \kt} \vb{\xi}_n(t).
\end{eqnarray}

Note that activity could have been implemented by making the beads active,
rather than the bonds. But, adding activity to the beads also requires
constraining the orientation of the active force, leading to additional
complexity (see Appendix~\ref{app:MeanSquareBondLength}). Our implementation of
activity is the most tractable for analytical computation, and successfully
captures the phenomenology of tangential driving as shown below.


\section{Analytical results}

Assuming a Rouse chain with a contour length much longer than the bond length
$b_0$, we take the continuous limit of Eq.~\eqref{eq:DiscreteRouseEOM} to obtain
\begin{equation} \label{eq:RouseEOMDimensional}
    \gamma \pdv{\vb{r}(n,t)}{t} = k \pdv[2]{\vb{r}(n,t)}{n}
    + \fa \pdv{\vb{r}(n,t)}{n}
    + \sqrt{2 \gamma \kt} \vb{\xi}(n, t)
\end{equation}
with the boundary conditions
\begin{equation} \label{eq:RouseEOMBoundaryConds}
    \pdv{\vb{r}(n,t)}{n}\eval_{n=0,N} = 0,
\end{equation}
which physically correspond to force-free boundary conditions. We
non-dimensionalize this equation by measuring time in units of $d\gamma/k$,
distance in units of $b_0 = \sqrt{d\kt/k}$, and energy in units of $\kt$.
Finally, we let $\alpha = \fa N b_0^2/2d\kt$ be a measure of activity in our
system. As such, $\alpha$ is a measure of the ratio of work performed by the
active force to the thermal energy. Our equation of motion now takes the form
\begin{equation} \label{eq:RouseEOM}
    \pdv{\tilde{\vb{r}}}{\tilde{t}}
    = d \pdv[2]{\tilde{\vb{r}}}{n} + 
        \frac{2 d \alpha}{N} \pdv{\tilde{\vb{r}}}{n} +
        \sqrt{2} \tilde{\vb{\xi}}(n,\tilde{t}).
\end{equation}


\subsection{Eigenfunction expansion}
The general solution of~\eqref{eq:RouseEOM} is
\begin{equation} \label{eq:EigenfunctionExpansion}
    \vb{r}(n, t) = \sum_{p=0}^\infty \vb{c}_p(t) \phi_p(n)
\end{equation}
where the $\phi_p(n)$ are the eigenfunctions
\begin{equation} \label{eq:Eigenfunctions}
    \phi_p(n) = A_p e^{-\alpha n / N} 
    \left[
        \cos (\frac{\omega_p n}{N}) + 
        \frac{\alpha}{\omega_p} \sin(\frac{\omega_p n}{N})
    \right]
\end{equation}
where $\omega_p = \pi p + i \alpha \delta_{p,0}$, and $A_p$ is a
normalization factor:
\begin{equation}
    \label{eq:NormalizationFactor}
    A_p^2 = \frac{2}{N} \begin{cases}
        \alpha e^{-\alpha} / 2 \sinh \alpha & p = 0, \\
        \pi^2 p^2 / (\pi^2 p^2 + \alpha^2) & p > 0.
    \end{cases}
\end{equation}
The decaying exponential in these eigenfunctions encodes the breaking of the
head-tail symmetry due to the active forces.

The $\phi_p$ are orthonormal with respect to the weight function $w(n) =
e^{2\alpha n / N}$; that is,
\begin{equation} \label{eq:EigenfunctionInnerProduct}
    \int_0^N \dd{n} w(n) \phi_p(n) \phi_q(n)
    = \delta_{pq}.
\end{equation}
We can check that in the limit $\alpha \to 0$, this reduces to the standard
cosine series, which is the correct set of eigenfunctions for the passive Rouse
chain~\cite{doi2007}.

Without loss of generality, we assume $\vb{r}(n, 0) = 0$, so that
\begin{equation} \label{eq:CpSolution}
    \vb{c}_p(t) = \int_0^t \dd{s} e^{-\lambda_p^2 (t - s)} \vb{\xi}_p(s),
\end{equation}
where
\begin{equation}
    \label{eq:TimeConstants}
    \lambda_p^2 = \frac{d}{N^2} \times 
        [\pi^2 p^2 + (1 - \delta_{p,0}) \alpha^2]
\end{equation}
and
\begin{equation}
    \label{eq:NoiseModes}
    \vb{\xi}_p(t) =
    \sqrt{2} \int_0^N \dd{n} w(n) \phi_p(n) \vb{\xi}(n, t).
\end{equation}
Unlike a passive Rouse chain, these noise modes are now correlated, so that
\begin{equation}
    \label{eq:NoiseModeMoments}
    \ev{\xi_p^\alpha(t)\xi_q^\beta(t')}
    = G_{pq} \delta_{\alpha \beta} \delta(t - t')
\end{equation}
where
\begin{equation} \label{eq:NoiseModeCorrelation}
    G_{pq} = 2 \int_0^N \dd{n} w(n)^2 \phi_p(n) \phi_q(n).
\end{equation}

We now use the eigenfuncton representation of the exact solution to compute the
center-of-mass diffusion coefficient and RMS end-to-end distance.


\subsection{Diffusion coefficient}

The center of mass $\vb{X}(t)$ of the chain is given by
\begin{equation} \label{eq:CenterOfMass}
    \vb{X}(t) = \frac{1}{N} \int_0^N \dd{n} \vb{r}(n, t)
    = \sum_{p=0}^\infty \vb{c}_p(t) \bar{\phi}_p
\end{equation}
where $\bar{\phi}_p = \int \dd{n} \phi_p(n) / N$ is the average value of
$\phi_p$ over the interval $n \in [0, N]$. From this, we compute the
mean square displacement (MSD) as
\begin{eqnarray}
    \label{eq:MeanSquareDisplacement}
    \text{MSD} &=& \ev{X(t)^2}
    \nonumber \\
    &=& \sum_{p,q} \bar{\phi}_p \bar{\phi}_q \ev{\vb{c}_p(t) \cdot \vb{c}_q(t)}
    \nonumber \\
    &=& d G_{00} \bar{\phi}_0^2 t + F(t),
\end{eqnarray}
where $F(t)$ is a function that contains only terms that are constant or decay
with time (see Appendix~\ref{app:MeanSquareDisplacement}). From this, we find
the diffusion coefficient to be
\begin{equation} \label{eq:DiffusionCoefficient}
    D(\alpha)
    = \lim_{t \to \infty} \frac{\text{MSD}}{2 d t}
    = \frac{1}{2} G_{00} \phi_0^2
    = D_0 \alpha \coth \alpha.
\end{equation}
where $D_0 = 1/N$ is the diffusion coefficient of a passive Rouse chain.
Notably, this gives the limiting behaviors
\begin{equation}
    \label{eq:DiffusionLimitingBehavior}
    D(\alpha) \propto \begin{cases}
        D_0 (1 + \alpha^2/3) & \alpha \ll 1, \\
        D_0 \alpha & \alpha \gg 1,
    \end{cases}
\end{equation}
This result shows that when the active work per bead is small compared to the
thermal energy, the diffusion coefficient grows with the square of activity,
which is reminiscent of the behavior of an active Brownian
particle~\cite{ghosh2015}. However, for $\alpha \gg 1$, i.e., when the active
work is large compared to the thermal fluctuations, the diffusion coefficient
grows linearly with activity.

At timescales shorter than the rotational relaxation time (discussed in the next
section), we expect to see evidence of the active driving through a nonlinear
growth in the MSD. After considering the possible effects of inertia, the MSD
takes the form
\[
    \textrm{MSD}(\delta t) \approx 2dD(\alpha) \delta t + B(\alpha) \delta t^2
\]
for $\delta t \ll 1$. Here, the coefficient $B(\alpha)$ describes the ballistic
motion of the filament at short times scales. Interestingly, we can show that
$B(\alpha)$ actually decays with increasing $\alpha$. This is discussed further
in Appendix~\ref{app:EffectOfInertia}.


\subsection{Conformational Dynamics}
There are two relevant parameters that encode the conformational dynamics of the
active polymer: the end-to-end length (or the radius of gyration, see
Appendix~\ref{app:RadiusOfGyration}), which captures its size, and the
relaxation time over which correlations in the end-to-end vector decay. We
compute each of these quantities here.

The end-to-end vector $\vb{L}$ is given by
\begin{equation}
    \label{eq:EndToEndVector}
    \vb{L}(t) = \vb{r}(N,t) - \vb{r}(0, t)
    = \sum_{p>0}^\infty \vb{c}_p(t) \Delta \phi_p,
\end{equation}
where $\Delta \phi_p = \phi_p(N) - \phi_p(0)$, and the $p = 0$ mode vanishes
since $\Delta \phi_0 = 0$. In the long-time limit, we obtain
\begin{eqnarray} \label{eq:EndToEndDistanceExpression}
    \ev{L^2}
    &=& \sum_{p,q>0} \Delta \phi_p \Delta \phi_q
    \lim_{t \to \infty} \ev{\vb{c}_p(t) \cdot \vb{c}_q(t)}
    \nonumber \\
    &=& d \sum_{p,q>0} G_{pq}
    \frac{\Delta \phi_p \Delta \phi_q}{\lambda_p^2 + \lambda_q^2}.
\end{eqnarray}
While this series representation is the exact result, it converges slowly and
does not lend itself to analytical approximation. We compute the sum numerically
(see Appendix~\ref{app:MeanSquareEndToEndDistance}) and find that
\begin{equation}
    \label{eq:EndToEndDistance}
    \ev{L^2} \approx N.
\end{equation}
That is, $\ev{L^2}$ is independent of the strength of the active force.

This result can be understood in the context of the microscopic equations of
motion given in Eq.~\eqref{eq:DiscreteRouseEOM} as follows. Since there are no
terms that lead to correlations in bond vector orientations in the model, we can
envision the polymer as being constructed of uncorrelated active rods. The
active forces exerted by these rods  cannot change their own lengths,
and so the overall length of the polymer is left unchanged. Interestingly, this
result does not necessarily hold if the beads are made active instead of the
bonds (see the appendix for more details). Further, we expect this result to be
modified in the context of the semiflexible polymer where orientational
correlations between the bonds can modify the end-to-end length, as discussed in
the subsequent sections.

Next, we compute the end-to-end vector autocorrelation function to obtain the
rotational relaxation time, $\tau_R$, using the approximation
\begin{equation}
    \label{eq:EndToEndCorrelationApprox}
    \ev{\vb{L}(t + \tau) \cdot \vb{L}(t)}
    \propto e^{-\tau/\tau_R}
\end{equation}
in the $t \to \infty$ limit. Using \eqref{eq:EndToEndVector}, we see that
\begin{eqnarray}
    \label{eq:RotationalRelaxation}
    \ev{\vb{L}(t + \tau) \cdot \vb{L}(t)}
    &=& \sum_{p,q>0} \Delta \phi_p \Delta \phi_q
    \ev{\vb{c}_p(t + \tau) \cdot \vb{c}_q(t)}
    \nonumber \\
    &=& d\sum_{p,q>0} G_{pq}
    \frac{\Delta \phi_p \Delta \phi_q}{\lambda_p^2 + \lambda_q^2}
    e^{-\lambda_p^2 \tau}.
\end{eqnarray}
As with Eq.~\eqref{eq:EndToEndDistanceExpression}, this sum cannot be computed
analytically. Assuming it can be approximated by the slowest decaying term
($\propto e^{-\lambda_1^2 t}$), we find the rotational relaxation time to be
\begin{equation}
    \label{eq:RelaxationTime}
    \tau_R = 1/\lambda_1^2 = \frac{\tau_R^0}{1 + \alpha^2/\pi^2},
\end{equation}
where $\tau_R^0 = N^2 / \pi^2 d$ is the relaxation time of a passive Rouse
filament. Activity therefore reduces the relaxation time. It is worth noting
that the approximation in Eq.~\eqref{eq:EndToEndCorrelationApprox} is limited in
that no individual term of Eq.~\eqref{eq:RotationalRelaxation} dominates. Though
Eq.~\eqref{eq:RelaxationTime} is the slowest relaxation time, it is not
necessarily the dominant one. See Appendix~\ref{app:RotationalRelaxationTime}
for more details.

\subsection{Mapping to a Semiflexible Filament}

Now, we generalize the above results to the case of a
semiflexible polymer. Consider a filament as a chain of $N$ beads connected via
inextensible bonds of length $b_0$, with rigidity encoded through the potential
\[
    H(\{\vb{r}_i\}) = \frac{1}{2} \kappa
        \sum_{i=1}^{N-2} \vu{t}_i \cdot \vu{t}_{i+1},
\]
where $\vb{t}_i = \vb{r}_{i+1} - \vb{r}_i$ and $\vu{t}_i = \vb{t}_i /
|\vb{t}_i|$. Activity is added in the same manner as in the case of the Rouse
filament. For simplicity, we will neglect excluded volume interactions in these
considerations; these will be incorporated in our computational model later.

Suppose our semiflexible filament is constructed of $N$ bonds with typical bond
length $b_0$. We can also view the filament as being constructed of $n$ rigid
segments of length $b = 2 \lp$. Then using $Nb_0 = n b$, we have
\[
    \alpha = \frac{\fa N b_0^2}{2 d \kt}
    = \frac{\fa n b^2}{2 d \kt} \frac{b_0}{b}
    = \frac{\tilde{\alpha} b_0}{2\lp}.
\]
That is, there is an effective activity $\tilde{\alpha}$ for a semiflexible
filament that is related to that of a simple Rouse chain through
\begin{equation} \label{eq:SemiflexibleActivity}
    \tilde{\alpha} = 2 \alpha \lp / b_0.
\end{equation}
We hypothesize that if we substitute this renormalized activity into the results
for the Rouse chain, they will generalize to the case of an active semiflexible
polymer. In particular, Eq.~\eqref{eq:DiffusionCoefficient} becomes
\begin{equation} \label{eq:ModifiedDiffusion}
    D(\alpha, \kappa) / D_0
    = (2 \alpha \lp / b_0) \coth (2 \alpha \lp / b_0),
\end{equation}
which implicitly depends on the stiffness $\kappa$ through $\lp$. We test this
hypothesis using numerical simulations in the next section.


\begin{figure}
    \centering
    \includegraphics{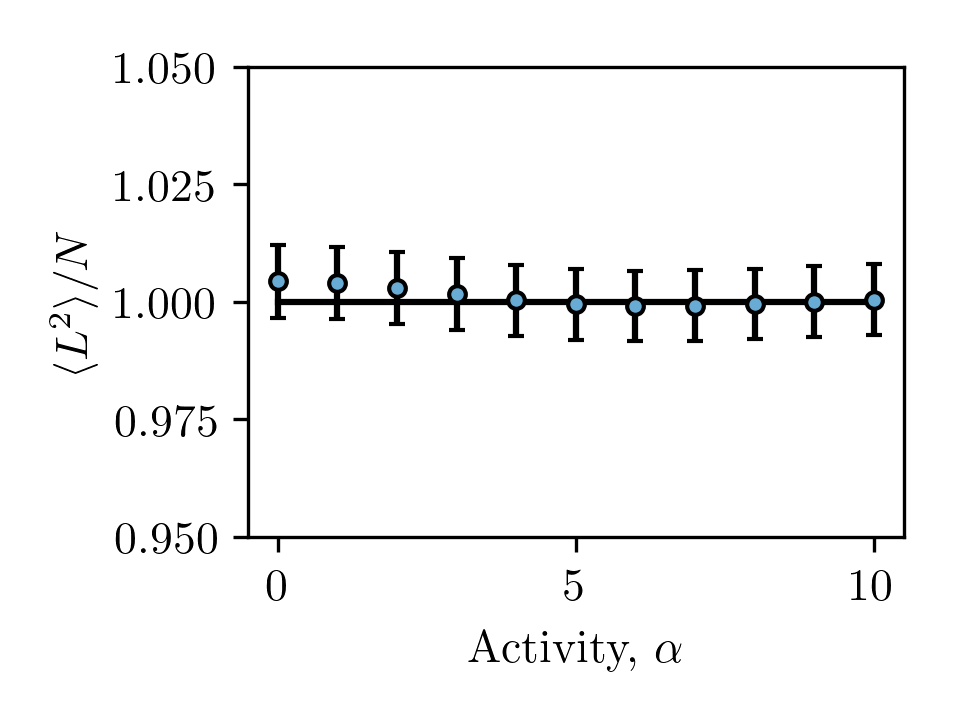}
    \caption{The mean square end-to-end length $\ev{L^2}$ normalized by the
    number of bonds $N$ is shown as a function of activity $\alpha$. The
    horizontal line is the predicted value based on
    Eq.~\eqref{eq:EndToEndDistance}, the symbols are results from computer
    simulations, and the error bars show the 95\% confidence interval. For all
    simulation results in this article, we used $N_\text{atoms}=51$ beads.}
    \label{fig:EndToEndLength}
\end{figure}

\begin{figure}
    \centering
    \includegraphics{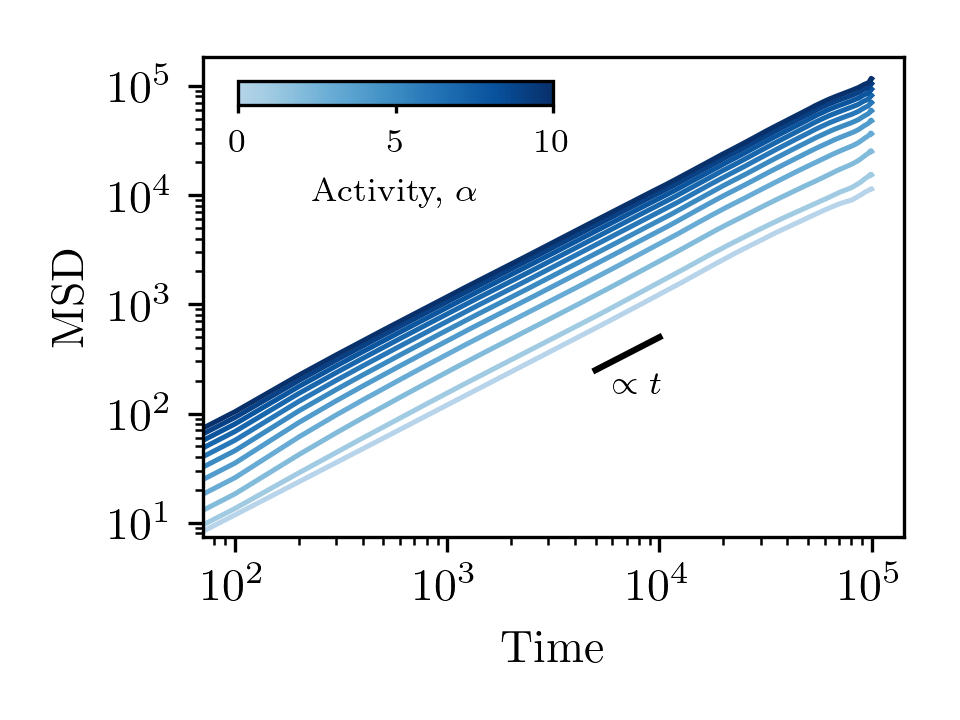}
    \includegraphics{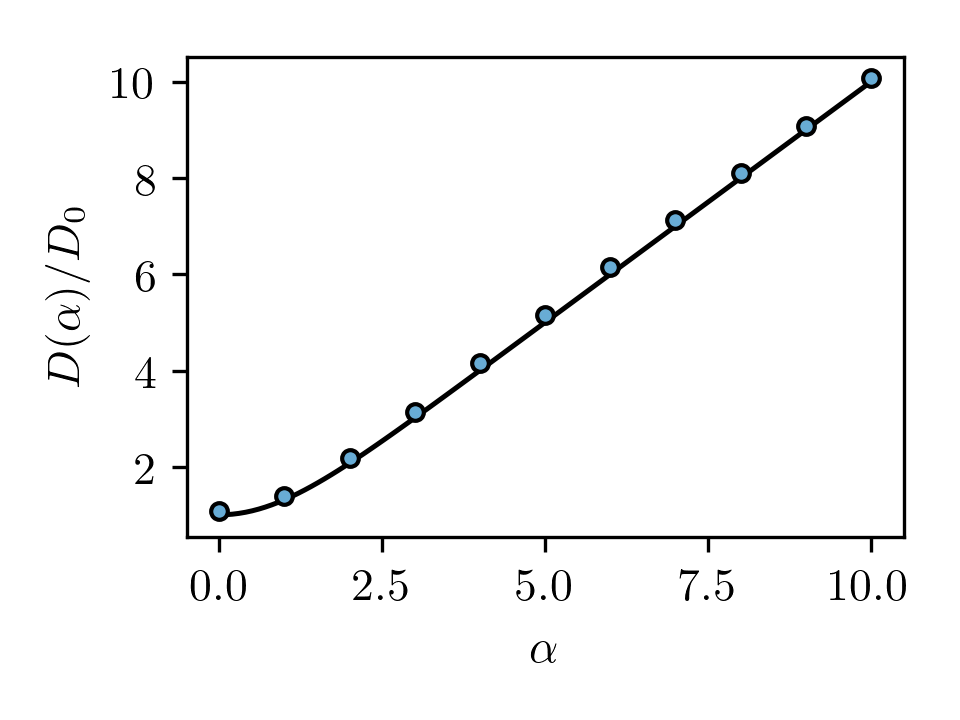}
    \caption{Dynamics of active Rouse chains. \textbf{Top:} The mean square
    displacement (MSD) for Rouse chains as a function of time for various
    activities, computed from simulation trajectories. For the range of times
    shown, all of the filaments exhibited purely diffusive motion, with higher
    activities leading to larger growth in the MSD with time. \textbf{Bottom:}
    The ratio of the active diffusion coefficient $D(\alpha)$ to the passive
    diffusion coefficient $D_0$ for a range of activities. The points are
    simulation data, and the line is the prediction from
    Eq.~\eqref{eq:DiffusionCoefficient}.}
    \label{fig:RouseDiffusionCoefficient}
\end{figure}

\begin{figure}
    \centering
    \includegraphics{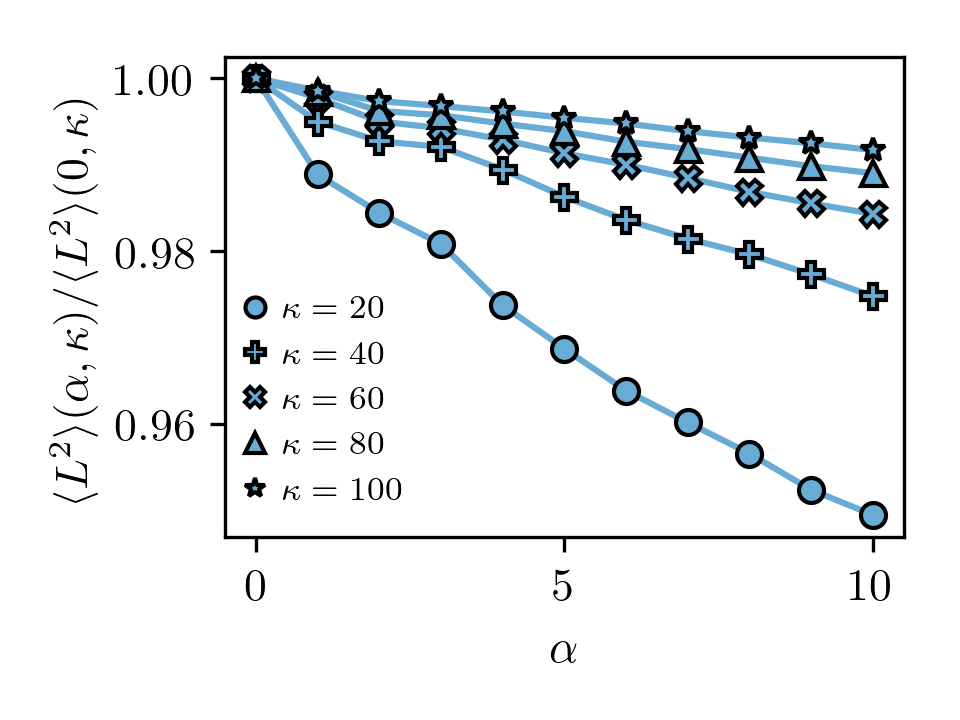}
    \includegraphics{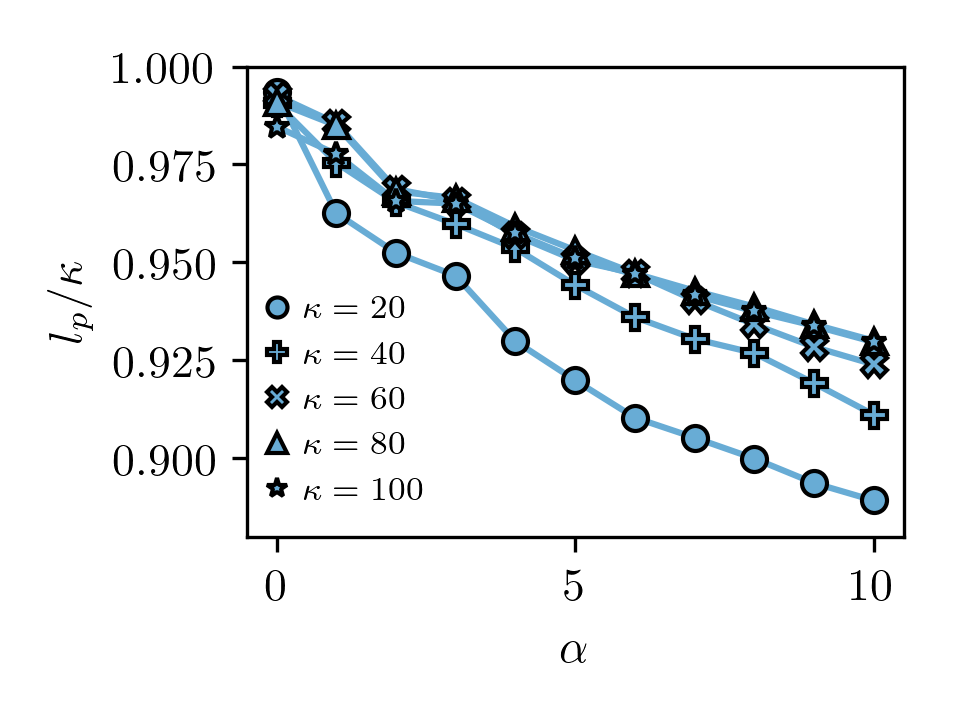}
    \caption{Configurational properties of active semiflexible filaments.
    \textbf{Top:} The mean-squared end-to-end length computed from simulation
    trajectories as a function of the activity $\alpha$ for various stiffnesses
    $\kappa$. The results are normalized by those of the passive ($\alpha = 0$)
    case. In general, we observe $\ev{L^2}$ decaying with $\alpha$, though this
    effect is weak over the range of activities tested. \textbf{Bottom:}
    Persistence length $\lp$ normalized by $\kappa$ as a function of activity.
    This plot more clearly shows  the reduction in $\lp$ with increasing
    activity, indicating a slight softening of the filament.}
    \label{fig:PersistenceLength}
\end{figure}

\begin{figure}
    \centering
    \includegraphics{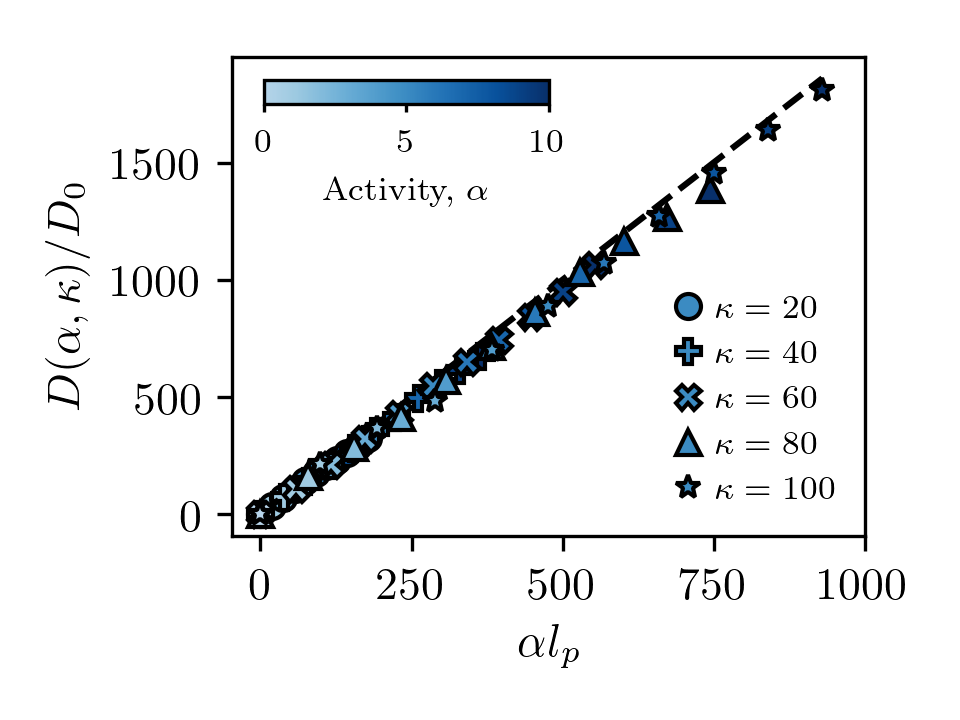}
    \includegraphics{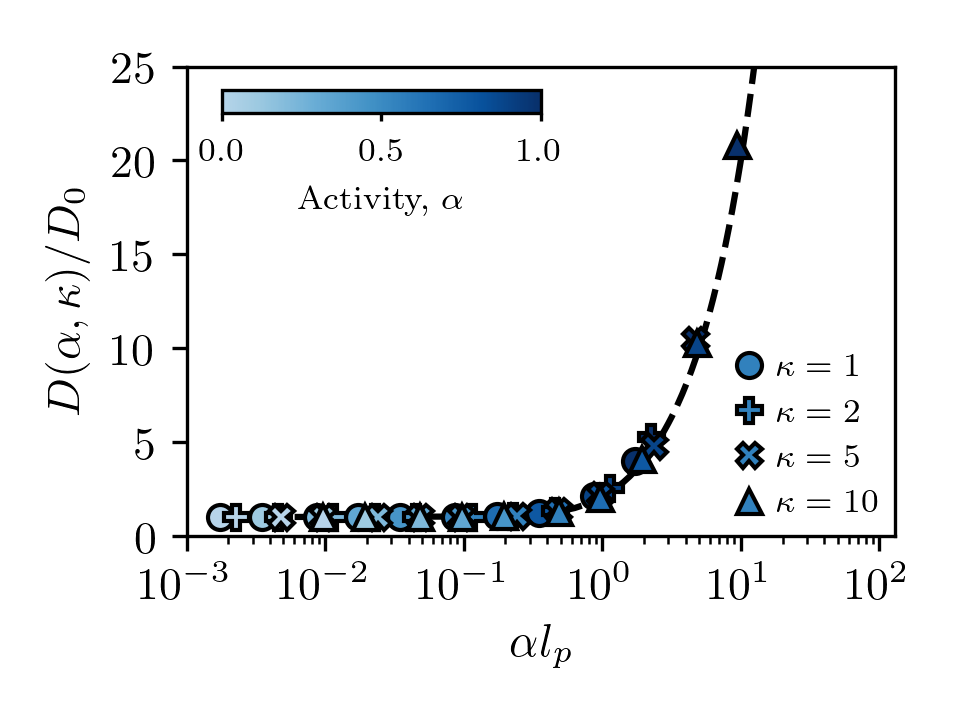}
    \caption{Dynamics of active semiflexible filaments. \textbf{Top:} Diffusion
    coefficients $D(\alpha, \kappa)$ for all simulation parameters as a function
    of $\alpha \lp$, the typical net active force exerted on a correlated
    segment of the filament of length $\lp$. All of the data lie along the line
    $D/D_0 \sim 2 \alpha \lp$ (dashed line). Notably, we see the same linear
    scaling of the diffusion coefficient with activity as with the active Rouse
    chain (see Fig.~\ref{fig:RouseDiffusionCoefficient}). \textbf{Bottom:}
    Diffusion coefficient measured at low $\alpha \lp$. The dashed line shows
    Eq.~\eqref{eq:ModifiedDiffusion} using the modified activity parameter from
    Eq.~\eqref{eq:SemiflexibleActivity}.}
    \label{fig:SemiflexibleDiffusionCoefficient}
\end{figure}

\section{Simulation results}

In this section, we perform numerical simulations to: a) validate the continuum
approximation of the Rouse chain used to obtain the analytical results, and b)
test the hypothesis of their generalization to semiflexible active polymers with
a renormalized activity parameter.

\subsection{Active Rouse Filaments}

To test the continuum limit approximation, we integrate the discrete equations
of motion of a Rouse chain, Eq.~\eqref{eq:DiscreteRouseEOM}, for a filament with
$N_\text{atoms} = 51$ atoms. As before, we non-dimensionalize by measuring
energy in units of $\kt$, time in units of $d \gamma / k$, and length in units
of $\sqrt{d k_B T / k}$. Additionally, $f_a = 2d\alpha / N$ where $0 \leq \alpha
\leq 10$ and $N = N_\text{atoms} - 1 = 50$ is the number of bonds. We use a
time step of $\Delta t = 10^{-3}$ and integrate for a total of $10^8$ steps.

We start by considering the steady-state mean square end-to-end length
$\ev{L^2}$. For a passive Rouse polymer, we know that $\ev{L^2} = N b_0^2$, and
we expect from Eq.~\eqref{eq:EndToEndDistance} that this will hold even for an
active filament. Indeed, we can see in Fig.~\ref{fig:EndToEndLength} that the
polymer size is independent of the strength of the active force. The same result
was found when testing chains with shorter lengths as well.

Next, note that based on Eq.~\eqref{eq:RelaxationTime}, the slowest relaxation
time occurs for a passive Rouse chain, for which $\tau_R \approx 10^2$ for the
units chosen here. Thus, for lag times $t > \tau_R$, the filament orientations
should decorrelate, and so the mean square displacement (MSD)  should grow
linearly in time. Fig.~\ref{fig:RouseDiffusionCoefficient} (top) shows the MSD
results computed from simulations, which exhibit diffusive motion for times $t
\gtrsim 10^2$ for all active force strengths. From these results, we can compute
the activity-dependent diffusion coefficient $D(\alpha)$
(Fig.~\ref{fig:RouseDiffusionCoefficient} (bottom)). Here, we observe close
agreement between the prediction of Eq.~\eqref{eq:DiffusionCoefficient} and the
simulation results. In particular, we see the expected linear scaling of the
diffusion coefficient with activity for $\alpha > 1$.


\subsection{Active Semiflexible Filaments}

To see if the results above are valid beyond the Rouse limit, we consider a more
realistic model that incorporates stiff bonds, resistance to bending, and
excluded volume interactions. We add these properties by including the potential
\begin{eqnarray}
    U(\{\vb{r}_n\})
    &=& \sum_{n=1}^{N-1} \frac{1}{2} k (|\vb{t}_n| - b_0)^2
    + \sum_{n=1}^{N-2} \kappa (1 - \vu{t}_n \cdot \vu{t}_{n+1})
    \nonumber \\
    &+& \sum_{i \neq j} U_\text{WCA}(|\vb{r}_j - \vb{r}_i|)
    \label{eq:SemiflexiblePotential}
\end{eqnarray}
where $\vb{t}_n = \vb{r}_{n+1} - \vb{r}_n$, $b_0$ is the preferred bond length,
and $k$ and $\kappa$ set the strength of the bond and angle potentials,
respectively. The potential $U_\text{WCA}(r)$ is a purely repulsive
Weeks-Chandler-Andersen potential~\cite{weeks1971}, defined as
\begin{equation}
    \label{eq:WCAPotential}
    U_\text{WCA}(r) = \begin{cases}
        4 \epsilon \left[
        \left(\dfrac{\sigma}{r}\right)^{12}
        - \left(\dfrac{\sigma}{r}\right)^6
        + \dfrac{1}{4}
        \right]
        & r \leq 2^{1/6}\sigma, \\
        0 & r > 2^{1/6}\sigma.
    \end{cases}
\end{equation}
The equation of motion for the $n$th bead of the semiflexible polymer is
therefore
\begin{equation}
    \label{eq:SemiflexibleEOM}
    \gamma \pdv{\vb{r}_n}{t}
    = -\pdv{U}{\vb{r}_n}
    + \fa \left(\frac{\vb{r}_{n+1} - \vb{r}_{n-1}}{2}\right)
    + \sqrt{2 \gamma \kt} \vb{\xi}_n(t).
\end{equation}
We choose $b_0 = \kt = 1$, $\sigma = b_0$, $\epsilon = \kt$, and $k = 200$,
and we vary $\kappa \in [20, 100]$.

We again start by investigating how activity affects the polymer size, in this
case looking at both the normalized mean-square end-to-end distance
$\ev{L^2}(\alpha,\kappa) / \ev{L^2}(0, \kappa)$ and the persistence length $\lp$
(see Fig.~\ref{fig:PersistenceLength}), which is computed by fitting the
tangent-tangent correlation function $C_{t}(m) = \ev{\vu{t}(n + m) \cdot
\vu{t}(n)}$ to the exponential $e^{-m b_0/\lp}$. In general, we find that
activity reduces the size of the polymer, but this effect is weak over the range
of activities tested. As such, we see that $\lp \approx \kappa$ for all
$\alpha$, as is expected for a passive semiflexible filament. The decrease in
persistence length with activity indicates that activity leads to a slight
``softening'' of the filament. This behavior has been studied in-depth in recent
work on polymers with directed active forces~\cite{anand2018, gupta2019,
bianco2018}.

We compute the diffusion coefficient $D(\alpha, \kappa)$ for semiflexible
polymers  the MSD results, and find linear scaling with the parameter $\alpha
\lp$ for $\alpha \lp \gtrsim 1$, as shown in
Fig.~\ref{fig:SemiflexibleDiffusionCoefficient}), with $D(\alpha, \kappa)
\approx 2 \alpha \lp$. To fully test the applicability of
Eq.~\eqref{eq:ModifiedDiffusion}, we performed additional simulations for
$\alpha \lp \ll 1$ and found excellent agreement between the measured and
predicted values over more than five orders of magnitude.


\section{Summary and Discussion}

In this paper, we consider a simple model for an active filament in the form of
a Rouse chain with an additional force acting along the tangent. By explicitly
solving the equation of motion, Eq.~\eqref{eq:RouseEOM}, we analytically compute
certain configurational and dynamical observables, in particular the MSD,
diffusion coefficient, and the end-to-end length as a function of the active
force strength. We find that the filament exhibits diffusive motion for times
larger than a rotational relaxation timescale, $\tau_R$, which decays rapidly
with activity (see Eq.~\eqref{eq:RelaxationTime}). This diffusive motion is
characterized by an activity-dependent diffusion coefficient that grows linearly
with $\alpha$ for $\alpha \gg 1$. This is in contrast to studies on passive
filaments in an active bath (see, for example,~\cite{eisenstecken2016,
osmanovic2017, chaki2019, ghosh2014}), which find the diffusion coefficient to
grow with the square of the active force strength. In general, studies of
polymers with configuration-independent active forces can be well described via
an ``effective temperature'', whereas configuration-dependent active forces, as
studied here, cannot be readily described in this manner.

We verify these analytical results by performing molecular dynamics simulations
of an active Rouse chain for a range of active force strengths. As evidenced by
Figs.~\ref{fig:EndToEndLength} and~\ref{fig:RouseDiffusionCoefficient}, we
observe excellent agreement between theory and simulation. To test whether the
active Rouse model results can be extended to more realistic polymer models, we
compare our analytical results to simulations of a semiflexible polymer with
excluded volume interactions and stiff bonds. We find that the persistence
length is weakly dependent on activity over the range of activities tested (as
shown in Fig.~\ref{fig:PersistenceLength}), with activity leading to a
``softening'' of the filament. This result is consistent with other recent
numerical studies on active polymers~\cite{anand2018, bianco2018, gupta2019}.

Additionally, we find that the diffusion coefficient $D(\alpha, \kappa)$ grows
linearly with a renormalized activity parameter, $\tilde{\alpha} = 2 \alpha \lp
/ b_0$. This can be explained by envisioning the semiflexible filament as a
Rouse chain with $n$ bonds of length $b$ equal to the Kuhn length $2\lp$, and
directly applying the results of Eq.~\eqref{eq:DiffusionCoefficient}. In
principle, this depends non-linearly on activity since the persistence length
$\lp$ is also activity-dependent. However, for small activities, $\lp$ is
approximately independent of $\alpha$, and so we recover the linear scaling of
the diffusion coefficient with activity.

Due to the simplicity of this model, it can be readily extended to incorporate
richer interactions at the single-filament level. For example, future work could
examine how attractive or solvent-mediated interactions affect the behavior of
these filaments. These analytical results for the effective activity on an
isolated filament can also serve as a starting point for understanding emergent
behaviors of dense systems containing many active filaments.


\begin{acknowledgments}
We acknowledge support from the Brandeis NSF \mbox{MRSEC}, Bioinspired Soft
Materials, DMR-1420382 (MP, AB, and MFH) and DMR-1855914 (MFH). Computational
resources were provided by the NSF through XSEDE computing resources (MCB090163)
and the Brandeis HPCC which is partially supported by the Brandeis MRSEC.
\end{acknowledgments}

\bibliographystyle{apsrev4-1}
\bibliography{bibliography}

\begin{thebibliography}{58}%
\makeatletter
\providecommand \@ifxundefined [1]{%
 \@ifx{#1\undefined}
}%
\providecommand \@ifnum [1]{%
 \ifnum #1\expandafter \@firstoftwo
 \else \expandafter \@secondoftwo
 \fi
}%
\providecommand \@ifx [1]{%
 \ifx #1\expandafter \@firstoftwo
 \else \expandafter \@secondoftwo
 \fi
}%
\providecommand \natexlab [1]{#1}%
\providecommand \enquote  [1]{``#1''}%
\providecommand \bibnamefont  [1]{#1}%
\providecommand \bibfnamefont [1]{#1}%
\providecommand \citenamefont [1]{#1}%
\providecommand \href@noop [0]{\@secondoftwo}%
\providecommand \href [0]{\begingroup \@sanitize@url \@href}%
\providecommand \@href[1]{\@@startlink{#1}\@@href}%
\providecommand \@@href[1]{\endgroup#1\@@endlink}%
\providecommand \@sanitize@url [0]{\catcode `\\12\catcode `\$12\catcode
  `\&12\catcode `\#12\catcode `\^12\catcode `\_12\catcode `\%12\relax}%
\providecommand \@@startlink[1]{}%
\providecommand \@@endlink[0]{}%
\providecommand \url  [0]{\begingroup\@sanitize@url \@url }%
\providecommand \@url [1]{\endgroup\@href {#1}{\urlprefix }}%
\providecommand \urlprefix  [0]{URL }%
\providecommand \Eprint [0]{\href }%
\providecommand \doibase [0]{http://dx.doi.org/}%
\providecommand \selectlanguage [0]{\@gobble}%
\providecommand \bibinfo  [0]{\@secondoftwo}%
\providecommand \bibfield  [0]{\@secondoftwo}%
\providecommand \translation [1]{[#1]}%
\providecommand \BibitemOpen [0]{}%
\providecommand \bibitemStop [0]{}%
\providecommand \bibitemNoStop [0]{.\EOS\space}%
\providecommand \EOS [0]{\spacefactor3000\relax}%
\providecommand \BibitemShut  [1]{\csname bibitem#1\endcsname}%
\let\auto@bib@innerbib\@empty
\bibitem [{\citenamefont {Ramaswamy}(2010)}]{ramaswamy2010}%
  \BibitemOpen
  \bibfield  {author} {\bibinfo {author} {\bibfnamefont {S.}~\bibnamefont
  {Ramaswamy}},\ }\href
  {http://www.annualreviews.org/doi/10.1146/annurev-conmatphys-070909-104101}
  {\bibfield  {journal} {\bibinfo  {journal} {Annu. Rev. Condens. Matter
  Phys.}\ }\textbf {\bibinfo {volume} {1}},\ \bibinfo {pages} {323} (\bibinfo
  {year} {2010})}\BibitemShut {NoStop}%
\bibitem [{\citenamefont {Marchetti}\ \emph {et~al.}(2013)\citenamefont
  {Marchetti}, \citenamefont {Joanny}, \citenamefont {Ramaswamy}, \citenamefont
  {Liverpool}, \citenamefont {Prost}, \citenamefont {Rao},\ and\ \citenamefont
  {Simha}}]{marchetti2013}%
  \BibitemOpen
  \bibfield  {author} {\bibinfo {author} {\bibfnamefont {M.~C.}\ \bibnamefont
  {Marchetti}}, \bibinfo {author} {\bibfnamefont {J.~F.}\ \bibnamefont
  {Joanny}}, \bibinfo {author} {\bibfnamefont {S.}~\bibnamefont {Ramaswamy}},
  \bibinfo {author} {\bibfnamefont {T.~B.}\ \bibnamefont {Liverpool}}, \bibinfo
  {author} {\bibfnamefont {J.}~\bibnamefont {Prost}}, \bibinfo {author}
  {\bibfnamefont {M.}~\bibnamefont {Rao}}, \ and\ \bibinfo {author}
  {\bibfnamefont {R.~A.}\ \bibnamefont {Simha}},\ }\href
  {https://link.aps.org/doi/10.1103/RevModPhys.85.1143} {\bibfield  {journal}
  {\bibinfo  {journal} {Rev. Mod. Phys.}\ }\textbf {\bibinfo {volume} {85}},\
  \bibinfo {pages} {1143} (\bibinfo {year} {2013})}\BibitemShut {NoStop}%
\bibitem [{\citenamefont {Sanchez}\ \emph {et~al.}(2012)\citenamefont
  {Sanchez}, \citenamefont {Chen}, \citenamefont {DeCamp}, \citenamefont
  {Heymann},\ and\ \citenamefont {Dogic}}]{sanchez2012}%
  \BibitemOpen
  \bibfield  {author} {\bibinfo {author} {\bibfnamefont {T.}~\bibnamefont
  {Sanchez}}, \bibinfo {author} {\bibfnamefont {D.~T.~N.}\ \bibnamefont
  {Chen}}, \bibinfo {author} {\bibfnamefont {S.~J.}\ \bibnamefont {DeCamp}},
  \bibinfo {author} {\bibfnamefont {M.}~\bibnamefont {Heymann}}, \ and\
  \bibinfo {author} {\bibfnamefont {Z.}~\bibnamefont {Dogic}},\ }\href
  {http://www.nature.com/articles/nature11591} {\bibfield  {journal} {\bibinfo
  {journal} {Nature}\ }\textbf {\bibinfo {volume} {491}},\ \bibinfo {pages}
  {431} (\bibinfo {year} {2012})}\BibitemShut {NoStop}%
\bibitem [{\citenamefont {Giomi}\ \emph {et~al.}(2013)\citenamefont {Giomi},
  \citenamefont {Bowick}, \citenamefont {Ma},\ and\ \citenamefont
  {Marchetti}}]{giomi2013}%
  \BibitemOpen
  \bibfield  {author} {\bibinfo {author} {\bibfnamefont {L.}~\bibnamefont
  {Giomi}}, \bibinfo {author} {\bibfnamefont {M.~J.}\ \bibnamefont {Bowick}},
  \bibinfo {author} {\bibfnamefont {X.}~\bibnamefont {Ma}}, \ and\ \bibinfo
  {author} {\bibfnamefont {M.~C.}\ \bibnamefont {Marchetti}},\ }\href
  {https://link.aps.org/doi/10.1103/PhysRevLett.110.228101} {\bibfield
  {journal} {\bibinfo  {journal} {Phys. Rev. Lett.}\ }\textbf {\bibinfo
  {volume} {110}},\ \bibinfo {pages} {228101} (\bibinfo {year}
  {2013})}\BibitemShut {NoStop}%
\bibitem [{\citenamefont {DeCamp}\ \emph {et~al.}(2015)\citenamefont {DeCamp},
  \citenamefont {Redner}, \citenamefont {Baskaran}, \citenamefont {Hagan},\
  and\ \citenamefont {Dogic}}]{decamp2015}%
  \BibitemOpen
  \bibfield  {author} {\bibinfo {author} {\bibfnamefont {S.~J.}\ \bibnamefont
  {DeCamp}}, \bibinfo {author} {\bibfnamefont {G.~S.}\ \bibnamefont {Redner}},
  \bibinfo {author} {\bibfnamefont {A.}~\bibnamefont {Baskaran}}, \bibinfo
  {author} {\bibfnamefont {M.~F.}\ \bibnamefont {Hagan}}, \ and\ \bibinfo
  {author} {\bibfnamefont {Z.}~\bibnamefont {Dogic}},\ }\href
  {http://www.nature.com/articles/nmat4387} {\bibfield  {journal} {\bibinfo
  {journal} {Nat. Mater.}\ }\textbf {\bibinfo {volume} {14}},\ \bibinfo {pages}
  {1110} (\bibinfo {year} {2015})}\BibitemShut {NoStop}%
\bibitem [{\citenamefont {Doostmohammadi}\ \emph {et~al.}(2018)\citenamefont
  {Doostmohammadi}, \citenamefont {Ign{\'{e}}s-Mullol}, \citenamefont
  {Yeomans},\ and\ \citenamefont {Sagu{\'{e}}s}}]{doostmohammadi2018}%
  \BibitemOpen
  \bibfield  {author} {\bibinfo {author} {\bibfnamefont {A.}~\bibnamefont
  {Doostmohammadi}}, \bibinfo {author} {\bibfnamefont {J.}~\bibnamefont
  {Ign{\'{e}}s-Mullol}}, \bibinfo {author} {\bibfnamefont {J.~M.}\ \bibnamefont
  {Yeomans}}, \ and\ \bibinfo {author} {\bibfnamefont {F.}~\bibnamefont
  {Sagu{\'{e}}s}},\ }\href {https://doi.org/10.1038/s41467-018-05666-8}
  {\bibfield  {journal} {\bibinfo  {journal} {Nat . Commun.}\ }\textbf
  {\bibinfo {volume} {9}} (\bibinfo {year} {2018})}\BibitemShut {NoStop}%
\bibitem [{\citenamefont {Thampi}\ \emph {et~al.}(2014)\citenamefont {Thampi},
  \citenamefont {Golestanian},\ and\ \citenamefont {Yeomans}}]{thampi2014}%
  \BibitemOpen
  \bibfield  {author} {\bibinfo {author} {\bibfnamefont {S.~P.}\ \bibnamefont
  {Thampi}}, \bibinfo {author} {\bibfnamefont {R.}~\bibnamefont {Golestanian}},
  \ and\ \bibinfo {author} {\bibfnamefont {J.~M.}\ \bibnamefont {Yeomans}},\
  }\href
  {http://stacks.iop.org/0295-5075/105/i=1/a=18001?key=crossref.cd21aad1b853e1671cdb9a8439a5c6b0}
  {\bibfield  {journal} {\bibinfo  {journal} {EPL}\ }\textbf {\bibinfo {volume}
  {105}},\ \bibinfo {pages} {18001} (\bibinfo {year} {2014})}\BibitemShut
  {NoStop}%
\bibitem [{\citenamefont {Redner}\ \emph {et~al.}(2013)\citenamefont {Redner},
  \citenamefont {Hagan},\ and\ \citenamefont {Baskaran}}]{redner2013}%
  \BibitemOpen
  \bibfield  {author} {\bibinfo {author} {\bibfnamefont {G.~S.}\ \bibnamefont
  {Redner}}, \bibinfo {author} {\bibfnamefont {M.~F.}\ \bibnamefont {Hagan}}, \
  and\ \bibinfo {author} {\bibfnamefont {A.}~\bibnamefont {Baskaran}},\ }\href
  {https://link.aps.org/doi/10.1103/PhysRevLett.110.055701} {\bibfield
  {journal} {\bibinfo  {journal} {Phys. Rev. Lett.}\ }\textbf {\bibinfo
  {volume} {110}},\ \bibinfo {pages} {055701} (\bibinfo {year}
  {2013})}\BibitemShut {NoStop}%
\bibitem [{\citenamefont {Bechinger}\ \emph {et~al.}(2016)\citenamefont
  {Bechinger}, \citenamefont {Di~Leonardo}, \citenamefont {Löwen},
  \citenamefont {Reichhardt}, \citenamefont {Volpe},\ and\ \citenamefont
  {Volpe}}]{bechinger2016}%
  \BibitemOpen
  \bibfield  {author} {\bibinfo {author} {\bibfnamefont {C.}~\bibnamefont
  {Bechinger}}, \bibinfo {author} {\bibfnamefont {R.}~\bibnamefont
  {Di~Leonardo}}, \bibinfo {author} {\bibfnamefont {H.}~\bibnamefont {Löwen}},
  \bibinfo {author} {\bibfnamefont {C.}~\bibnamefont {Reichhardt}}, \bibinfo
  {author} {\bibfnamefont {G.}~\bibnamefont {Volpe}}, \ and\ \bibinfo {author}
  {\bibfnamefont {G.}~\bibnamefont {Volpe}},\ }\href
  {https://link.aps.org/doi/10.1103/RevModPhys.88.045006} {\bibfield  {journal}
  {\bibinfo  {journal} {Rev. Mod. Phys.}\ }\textbf {\bibinfo {volume} {88}},\
  \bibinfo {pages} {045006} (\bibinfo {year} {2016})}\BibitemShut {NoStop}%
\bibitem [{\citenamefont {Chelakkot}\ \emph {et~al.}(2014)\citenamefont
  {Chelakkot}, \citenamefont {Gopinath}, \citenamefont {Mahadevan},\ and\
  \citenamefont {Hagan}}]{chelakkot2014}%
  \BibitemOpen
  \bibfield  {author} {\bibinfo {author} {\bibfnamefont {R.}~\bibnamefont
  {Chelakkot}}, \bibinfo {author} {\bibfnamefont {A.}~\bibnamefont {Gopinath}},
  \bibinfo {author} {\bibfnamefont {L.}~\bibnamefont {Mahadevan}}, \ and\
  \bibinfo {author} {\bibfnamefont {M.~F.}\ \bibnamefont {Hagan}},\ }\href
  {https://royalsocietypublishing.org/doi/abs/10.1098/rsif.2013.0884}
  {\bibfield  {journal} {\bibinfo  {journal} {J. R. Soc., Interface}\ }\textbf
  {\bibinfo {volume} {11}},\ \bibinfo {pages} {20130884} (\bibinfo {year}
  {2014})}\BibitemShut {NoStop}%
\bibitem [{\citenamefont {Elgeti}\ \emph {et~al.}(2015)\citenamefont {Elgeti},
  \citenamefont {Winkler},\ and\ \citenamefont {Gompper}}]{elgeti2015}%
  \BibitemOpen
  \bibfield  {author} {\bibinfo {author} {\bibfnamefont {J.}~\bibnamefont
  {Elgeti}}, \bibinfo {author} {\bibfnamefont {R.~G.}\ \bibnamefont {Winkler}},
  \ and\ \bibinfo {author} {\bibfnamefont {G.}~\bibnamefont {Gompper}},\ }\href
  {http://stacks.iop.org/0034-4885/78/i=5/a=056601?key=crossref.098e1dfca7db8f1f68a5a86fca9219b5}
  {\bibfield  {journal} {\bibinfo  {journal} {Rep. Prog. Phys.}\ }\textbf
  {\bibinfo {volume} {78}},\ \bibinfo {pages} {056601} (\bibinfo {year}
  {2015})}\BibitemShut {NoStop}%
\bibitem [{\citenamefont {Bronstein}\ \emph {et~al.}(2009)\citenamefont
  {Bronstein}, \citenamefont {Israel}, \citenamefont {Kepten}, \citenamefont
  {Mai}, \citenamefont {Shav-Tal}, \citenamefont {Barkai},\ and\ \citenamefont
  {Garini}}]{bronstein2009}%
  \BibitemOpen
  \bibfield  {author} {\bibinfo {author} {\bibfnamefont {I.}~\bibnamefont
  {Bronstein}}, \bibinfo {author} {\bibfnamefont {Y.}~\bibnamefont {Israel}},
  \bibinfo {author} {\bibfnamefont {E.}~\bibnamefont {Kepten}}, \bibinfo
  {author} {\bibfnamefont {S.}~\bibnamefont {Mai}}, \bibinfo {author}
  {\bibfnamefont {Y.}~\bibnamefont {Shav-Tal}}, \bibinfo {author}
  {\bibfnamefont {E.}~\bibnamefont {Barkai}}, \ and\ \bibinfo {author}
  {\bibfnamefont {Y.}~\bibnamefont {Garini}},\ }\href
  {https://doi.org/10.1103/physrevlett.103.018102} {\bibfield  {journal}
  {\bibinfo  {journal} {Phys. Rev. Lett.}\ }\textbf {\bibinfo {volume} {103}}
  (\bibinfo {year} {2009})}\BibitemShut {NoStop}%
\bibitem [{\citenamefont {Bronshtein}\ \emph {et~al.}(2015)\citenamefont
  {Bronshtein}, \citenamefont {Kepten}, \citenamefont {Kanter}, \citenamefont
  {Berezin}, \citenamefont {Lindner}, \citenamefont {Redwood}, \citenamefont
  {Mai}, \citenamefont {Gonzalo}, \citenamefont {Foisner}, \citenamefont
  {Shav-Tal},\ and\ \citenamefont {Garini}}]{bronshtein2015}%
  \BibitemOpen
  \bibfield  {author} {\bibinfo {author} {\bibfnamefont {I.}~\bibnamefont
  {Bronshtein}}, \bibinfo {author} {\bibfnamefont {E.}~\bibnamefont {Kepten}},
  \bibinfo {author} {\bibfnamefont {I.}~\bibnamefont {Kanter}}, \bibinfo
  {author} {\bibfnamefont {S.}~\bibnamefont {Berezin}}, \bibinfo {author}
  {\bibfnamefont {M.}~\bibnamefont {Lindner}}, \bibinfo {author} {\bibfnamefont
  {A.~B.}\ \bibnamefont {Redwood}}, \bibinfo {author} {\bibfnamefont
  {S.}~\bibnamefont {Mai}}, \bibinfo {author} {\bibfnamefont {S.}~\bibnamefont
  {Gonzalo}}, \bibinfo {author} {\bibfnamefont {R.}~\bibnamefont {Foisner}},
  \bibinfo {author} {\bibfnamefont {Y.}~\bibnamefont {Shav-Tal}}, \ and\
  \bibinfo {author} {\bibfnamefont {Y.}~\bibnamefont {Garini}},\ }\href
  {http://www.nature.com/articles/ncomms9044} {\bibfield  {journal} {\bibinfo
  {journal} {Nat . Commun.}\ }\textbf {\bibinfo {volume} {6}},\ \bibinfo
  {pages} {8044} (\bibinfo {year} {2015})}\BibitemShut {NoStop}%
\bibitem [{\citenamefont {Cabal}\ \emph {et~al.}(2006)\citenamefont {Cabal},
  \citenamefont {Genovesio}, \citenamefont {Rodriguez-Navarro}, \citenamefont
  {Zimmer}, \citenamefont {Gadal}, \citenamefont {Lesne}, \citenamefont {Buc},
  \citenamefont {Feuerbach-Fournier}, \citenamefont {Olivo-Marin},
  \citenamefont {Hurt},\ and\ \citenamefont {Nehrbass}}]{cabal2006}%
  \BibitemOpen
  \bibfield  {author} {\bibinfo {author} {\bibfnamefont {G.~G.}\ \bibnamefont
  {Cabal}}, \bibinfo {author} {\bibfnamefont {A.}~\bibnamefont {Genovesio}},
  \bibinfo {author} {\bibfnamefont {S.}~\bibnamefont {Rodriguez-Navarro}},
  \bibinfo {author} {\bibfnamefont {C.}~\bibnamefont {Zimmer}}, \bibinfo
  {author} {\bibfnamefont {O.}~\bibnamefont {Gadal}}, \bibinfo {author}
  {\bibfnamefont {A.}~\bibnamefont {Lesne}}, \bibinfo {author} {\bibfnamefont
  {H.}~\bibnamefont {Buc}}, \bibinfo {author} {\bibfnamefont {F.}~\bibnamefont
  {Feuerbach-Fournier}}, \bibinfo {author} {\bibfnamefont {J.-C.}\ \bibnamefont
  {Olivo-Marin}}, \bibinfo {author} {\bibfnamefont {E.~C.}\ \bibnamefont
  {Hurt}}, \ and\ \bibinfo {author} {\bibfnamefont {U.}~\bibnamefont
  {Nehrbass}},\ }\href {https://doi.org/10.1038/nature04752} {\bibfield
  {journal} {\bibinfo  {journal} {Nature}\ }\textbf {\bibinfo {volume} {441}},\
  \bibinfo {pages} {770} (\bibinfo {year} {2006})}\BibitemShut {NoStop}%
\bibitem [{\citenamefont {Zidovska}\ \emph {et~al.}(2013)\citenamefont
  {Zidovska}, \citenamefont {Weitz},\ and\ \citenamefont
  {Mitchison}}]{zidovska2013}%
  \BibitemOpen
  \bibfield  {author} {\bibinfo {author} {\bibfnamefont {A.}~\bibnamefont
  {Zidovska}}, \bibinfo {author} {\bibfnamefont {D.~A.}\ \bibnamefont {Weitz}},
  \ and\ \bibinfo {author} {\bibfnamefont {T.~J.}\ \bibnamefont {Mitchison}},\
  }\href {http://www.pnas.org/cgi/doi/10.1073/pnas.1220313110} {\bibfield
  {journal} {\bibinfo  {journal} {Proc. Natl. Acad. Sci.}\ }\textbf {\bibinfo
  {volume} {110}},\ \bibinfo {pages} {15555} (\bibinfo {year}
  {2013})}\BibitemShut {NoStop}%
\bibitem [{\citenamefont {Ganai}\ \emph {et~al.}(2014)\citenamefont {Ganai},
  \citenamefont {Sengupta},\ and\ \citenamefont {Menon}}]{ganai2014}%
  \BibitemOpen
  \bibfield  {author} {\bibinfo {author} {\bibfnamefont {N.}~\bibnamefont
  {Ganai}}, \bibinfo {author} {\bibfnamefont {S.}~\bibnamefont {Sengupta}}, \
  and\ \bibinfo {author} {\bibfnamefont {G.~I.}\ \bibnamefont {Menon}},\ }\href
  {https://doi.org/10.1093/nar/gkt1417} {\bibfield  {journal} {\bibinfo
  {journal} {Nucleic Acids Res.}\ }\textbf {\bibinfo {volume} {42}},\ \bibinfo
  {pages} {4145} (\bibinfo {year} {2014})}\BibitemShut {NoStop}%
\bibitem [{\citenamefont {Haddad}\ \emph {et~al.}(2017)\citenamefont {Haddad},
  \citenamefont {Jost},\ and\ \citenamefont
  {Vaillant}}]{haddad2017perspectives}%
  \BibitemOpen
  \bibfield  {author} {\bibinfo {author} {\bibfnamefont {N.}~\bibnamefont
  {Haddad}}, \bibinfo {author} {\bibfnamefont {D.}~\bibnamefont {Jost}}, \ and\
  \bibinfo {author} {\bibfnamefont {C.}~\bibnamefont {Vaillant}},\ }\href
  {https://doi.org/10.1007/s10577-016-9548-2} {\bibfield  {journal} {\bibinfo
  {journal} {Chromosome Research}\ }\textbf {\bibinfo {volume} {25}},\ \bibinfo
  {pages} {35} (\bibinfo {year} {2017})}\BibitemShut {NoStop}%
\bibitem [{\citenamefont {Brangwynne}\ \emph {et~al.}(2008)\citenamefont
  {Brangwynne}, \citenamefont {Koenderink}, \citenamefont {MacKintosh},\ and\
  \citenamefont {Weitz}}]{brangwynne2008}%
  \BibitemOpen
  \bibfield  {author} {\bibinfo {author} {\bibfnamefont {C.~P.}\ \bibnamefont
  {Brangwynne}}, \bibinfo {author} {\bibfnamefont {G.~H.}\ \bibnamefont
  {Koenderink}}, \bibinfo {author} {\bibfnamefont {F.~C.}\ \bibnamefont
  {MacKintosh}}, \ and\ \bibinfo {author} {\bibfnamefont {D.~A.}\ \bibnamefont
  {Weitz}},\ }\href {https://link.aps.org/doi/10.1103/PhysRevLett.100.118104}
  {\bibfield  {journal} {\bibinfo  {journal} {Phys. Rev. Lett.}\ }\textbf
  {\bibinfo {volume} {100}},\ \bibinfo {pages} {118104} (\bibinfo {year}
  {2008})}\BibitemShut {NoStop}%
\bibitem [{\citenamefont {Mizuno}\ \emph {et~al.}(2007)\citenamefont {Mizuno},
  \citenamefont {Tardin}, \citenamefont {Schmidt},\ and\ \citenamefont
  {MacKintosh}}]{mizuno2007}%
  \BibitemOpen
  \bibfield  {author} {\bibinfo {author} {\bibfnamefont {D.}~\bibnamefont
  {Mizuno}}, \bibinfo {author} {\bibfnamefont {C.}~\bibnamefont {Tardin}},
  \bibinfo {author} {\bibfnamefont {C.~F.}\ \bibnamefont {Schmidt}}, \ and\
  \bibinfo {author} {\bibfnamefont {F.~C.}\ \bibnamefont {MacKintosh}},\ }\href
  {http://www.sciencemag.org/cgi/doi/10.1126/science.1134404} {\bibfield
  {journal} {\bibinfo  {journal} {Science}\ }\textbf {\bibinfo {volume}
  {315}},\ \bibinfo {pages} {370} (\bibinfo {year} {2007})}\BibitemShut
  {NoStop}%
\bibitem [{\citenamefont {Le~Goff}\ \emph {et~al.}(2001)\citenamefont
  {Le~Goff}, \citenamefont {Amblard},\ and\ \citenamefont
  {Furst}}]{le_goff2001}%
  \BibitemOpen
  \bibfield  {author} {\bibinfo {author} {\bibfnamefont {L.}~\bibnamefont
  {Le~Goff}}, \bibinfo {author} {\bibfnamefont {F.}~\bibnamefont {Amblard}}, \
  and\ \bibinfo {author} {\bibfnamefont {E.~M.}\ \bibnamefont {Furst}},\ }\href
  {https://link.aps.org/doi/10.1103/PhysRevLett.88.018101} {\bibfield
  {journal} {\bibinfo  {journal} {Phys. Rev. Lett.}\ }\textbf {\bibinfo
  {volume} {88}},\ \bibinfo {pages} {018101} (\bibinfo {year}
  {2001})}\BibitemShut {NoStop}%
\bibitem [{\citenamefont {Prathyusha}\ \emph {et~al.}(2018)\citenamefont
  {Prathyusha}, \citenamefont {Henkes},\ and\ \citenamefont
  {Sknepnek}}]{prathyusha2018}%
  \BibitemOpen
  \bibfield  {author} {\bibinfo {author} {\bibfnamefont {K.~R.}\ \bibnamefont
  {Prathyusha}}, \bibinfo {author} {\bibfnamefont {S.}~\bibnamefont {Henkes}},
  \ and\ \bibinfo {author} {\bibfnamefont {R.}~\bibnamefont {Sknepnek}},\
  }\href {https://link.aps.org/doi/10.1103/PhysRevE.97.022606} {\bibfield
  {journal} {\bibinfo  {journal} {Phys. Rev. E}\ }\textbf {\bibinfo {volume}
  {97}},\ \bibinfo {pages} {022606} (\bibinfo {year} {2018})}\BibitemShut
  {NoStop}%
\bibitem [{\citenamefont {Weber}\ \emph {et~al.}(2015)\citenamefont {Weber},
  \citenamefont {Suzuki}, \citenamefont {Schaller}, \citenamefont {Aranson},
  \citenamefont {Bausch},\ and\ \citenamefont {Frey}}]{weber2015}%
  \BibitemOpen
  \bibfield  {author} {\bibinfo {author} {\bibfnamefont {C.~A.}\ \bibnamefont
  {Weber}}, \bibinfo {author} {\bibfnamefont {R.}~\bibnamefont {Suzuki}},
  \bibinfo {author} {\bibfnamefont {V.}~\bibnamefont {Schaller}}, \bibinfo
  {author} {\bibfnamefont {I.~S.}\ \bibnamefont {Aranson}}, \bibinfo {author}
  {\bibfnamefont {A.~R.}\ \bibnamefont {Bausch}}, \ and\ \bibinfo {author}
  {\bibfnamefont {E.}~\bibnamefont {Frey}},\ }\href
  {http://www.pnas.org/lookup/doi/10.1073/pnas.1421322112} {\bibfield
  {journal} {\bibinfo  {journal} {Proc. Natl. Acad. Sci.}\ }\textbf {\bibinfo
  {volume} {112}},\ \bibinfo {pages} {10703} (\bibinfo {year}
  {2015})}\BibitemShut {NoStop}%
\bibitem [{\citenamefont {Humphrey}\ \emph {et~al.}(2002)\citenamefont
  {Humphrey}, \citenamefont {Duggan}, \citenamefont {Saha}, \citenamefont
  {Smith},\ and\ \citenamefont {Käs}}]{humphrey2002}%
  \BibitemOpen
  \bibfield  {author} {\bibinfo {author} {\bibfnamefont {D.}~\bibnamefont
  {Humphrey}}, \bibinfo {author} {\bibfnamefont {C.}~\bibnamefont {Duggan}},
  \bibinfo {author} {\bibfnamefont {D.}~\bibnamefont {Saha}}, \bibinfo {author}
  {\bibfnamefont {D.}~\bibnamefont {Smith}}, \ and\ \bibinfo {author}
  {\bibfnamefont {J.}~\bibnamefont {Käs}},\ }\href
  {http://www.nature.com/articles/416413a} {\bibfield  {journal} {\bibinfo
  {journal} {Nature}\ }\textbf {\bibinfo {volume} {416}},\ \bibinfo {pages}
  {413} (\bibinfo {year} {2002})}\BibitemShut {NoStop}%
\bibitem [{\citenamefont {Sakaue}\ and\ \citenamefont
  {Saito}(2017)}]{sakaue2017}%
  \BibitemOpen
  \bibfield  {author} {\bibinfo {author} {\bibfnamefont {T.}~\bibnamefont
  {Sakaue}}\ and\ \bibinfo {author} {\bibfnamefont {T.}~\bibnamefont {Saito}},\
  }\href {http://xlink.rsc.org/?DOI=C6SM00775A} {\bibfield  {journal} {\bibinfo
   {journal} {Soft Matter}\ }\textbf {\bibinfo {volume} {13}},\ \bibinfo
  {pages} {81} (\bibinfo {year} {2017})}\BibitemShut {NoStop}%
\bibitem [{\citenamefont {Suzuki}\ and\ \citenamefont
  {Bausch}(2017)}]{suzuki2017}%
  \BibitemOpen
  \bibfield  {author} {\bibinfo {author} {\bibfnamefont {R.}~\bibnamefont
  {Suzuki}}\ and\ \bibinfo {author} {\bibfnamefont {A.~R.}\ \bibnamefont
  {Bausch}},\ }\href {http://www.nature.com/articles/s41467-017-00035-3}
  {\bibfield  {journal} {\bibinfo  {journal} {Nat . Commun.}\ }\textbf
  {\bibinfo {volume} {8}},\ \bibinfo {pages} {41} (\bibinfo {year}
  {2017})}\BibitemShut {NoStop}%
\bibitem [{\citenamefont {Sokolov}\ \emph {et~al.}(2007)\citenamefont
  {Sokolov}, \citenamefont {Aranson}, \citenamefont {Kessler},\ and\
  \citenamefont {Goldstein}}]{sokolov2007}%
  \BibitemOpen
  \bibfield  {author} {\bibinfo {author} {\bibfnamefont {A.}~\bibnamefont
  {Sokolov}}, \bibinfo {author} {\bibfnamefont {I.~S.}\ \bibnamefont
  {Aranson}}, \bibinfo {author} {\bibfnamefont {J.~O.}\ \bibnamefont
  {Kessler}}, \ and\ \bibinfo {author} {\bibfnamefont {R.~E.}\ \bibnamefont
  {Goldstein}},\ }\href
  {https://link.aps.org/doi/10.1103/PhysRevLett.98.158102} {\bibfield
  {journal} {\bibinfo  {journal} {Phys. Rev. Lett.}\ }\textbf {\bibinfo
  {volume} {98}},\ \bibinfo {pages} {158102} (\bibinfo {year}
  {2007})}\BibitemShut {NoStop}%
\bibitem [{\citenamefont {Saintillan}\ \emph {et~al.}(2018)\citenamefont
  {Saintillan}, \citenamefont {Shelley},\ and\ \citenamefont
  {Zidovska}}]{saintillan2018}%
  \BibitemOpen
  \bibfield  {author} {\bibinfo {author} {\bibfnamefont {D.}~\bibnamefont
  {Saintillan}}, \bibinfo {author} {\bibfnamefont {M.~J.}\ \bibnamefont
  {Shelley}}, \ and\ \bibinfo {author} {\bibfnamefont {A.}~\bibnamefont
  {Zidovska}},\ }\href {https://doi.org/10.1073/pnas.1807073115} {\bibfield
  {journal} {\bibinfo  {journal} {Proc. Natl. Acad. Sci.}\ }\textbf {\bibinfo
  {volume} {115}},\ \bibinfo {pages} {11442} (\bibinfo {year}
  {2018})}\BibitemShut {NoStop}%
\bibitem [{\citenamefont {Gupta}\ \emph {et~al.}(2019)\citenamefont {Gupta},
  \citenamefont {Chaudhuri},\ and\ \citenamefont {Chaudhuri}}]{gupta2019}%
  \BibitemOpen
  \bibfield  {author} {\bibinfo {author} {\bibfnamefont {N.}~\bibnamefont
  {Gupta}}, \bibinfo {author} {\bibfnamefont {A.}~\bibnamefont {Chaudhuri}}, \
  and\ \bibinfo {author} {\bibfnamefont {D.}~\bibnamefont {Chaudhuri}},\ }\href
  {https://doi.org/10.1103/physreve.99.042405} {\bibfield  {journal} {\bibinfo
  {journal} {Phys. Rev. E}\ }\textbf {\bibinfo {volume} {99}} (\bibinfo {year}
  {2019})}\BibitemShut {NoStop}%
\bibitem [{\citenamefont {Isele-Holder}\ \emph {et~al.}(2015)\citenamefont
  {Isele-Holder}, \citenamefont {Elgeti},\ and\ \citenamefont
  {Gompper}}]{isele-holder2015}%
  \BibitemOpen
  \bibfield  {author} {\bibinfo {author} {\bibfnamefont {R.~E.}\ \bibnamefont
  {Isele-Holder}}, \bibinfo {author} {\bibfnamefont {J.}~\bibnamefont
  {Elgeti}}, \ and\ \bibinfo {author} {\bibfnamefont {G.}~\bibnamefont
  {Gompper}},\ }\href {http://xlink.rsc.org/?DOI=C5SM01683E} {\bibfield
  {journal} {\bibinfo  {journal} {Soft Matter}\ }\textbf {\bibinfo {volume}
  {11}},\ \bibinfo {pages} {7181} (\bibinfo {year} {2015})}\BibitemShut
  {NoStop}%
\bibitem [{\citenamefont {ten Hagen}\ \emph {et~al.}(2011)\citenamefont {ten
  Hagen}, \citenamefont {van Teeffelen},\ and\ \citenamefont
  {L{\"o}wen}}]{ten2011brownian}%
  \BibitemOpen
  \bibfield  {author} {\bibinfo {author} {\bibfnamefont {B.}~\bibnamefont {ten
  Hagen}}, \bibinfo {author} {\bibfnamefont {S.}~\bibnamefont {van Teeffelen}},
  \ and\ \bibinfo {author} {\bibfnamefont {H.}~\bibnamefont {L{\"o}wen}},\
  }\href {http://dx.doi.org/10.1088/0953-8984/23/19/194119} {\bibfield
  {journal} {\bibinfo  {journal} {Journal of Physics: Condensed Matter}\
  }\textbf {\bibinfo {volume} {23}},\ \bibinfo {pages} {194119} (\bibinfo
  {year} {2011})}\BibitemShut {NoStop}%
\bibitem [{\citenamefont {Angelani}\ \emph {et~al.}(2011)\citenamefont
  {Angelani}, \citenamefont {Costanzo},\ and\ \citenamefont
  {Di~Leonardo}}]{angelani2011active}%
  \BibitemOpen
  \bibfield  {author} {\bibinfo {author} {\bibfnamefont {L.}~\bibnamefont
  {Angelani}}, \bibinfo {author} {\bibfnamefont {A.}~\bibnamefont {Costanzo}},
  \ and\ \bibinfo {author} {\bibfnamefont {R.}~\bibnamefont {Di~Leonardo}},\
  }\href {https://doi.org/10.1209/0295-5075/96/68002} {\bibfield  {journal}
  {\bibinfo  {journal} {EPL (Europhysics Letters)}\ }\textbf {\bibinfo {volume}
  {96}},\ \bibinfo {pages} {68002} (\bibinfo {year} {2011})}\BibitemShut
  {NoStop}%
\bibitem [{\citenamefont {Ai}\ \emph {et~al.}(2013)\citenamefont {Ai},
  \citenamefont {Chen}, \citenamefont {He}, \citenamefont {Li},\ and\
  \citenamefont {Zhong}}]{ai2013rectification}%
  \BibitemOpen
  \bibfield  {author} {\bibinfo {author} {\bibfnamefont {B.-q.}\ \bibnamefont
  {Ai}}, \bibinfo {author} {\bibfnamefont {Q.-y.}\ \bibnamefont {Chen}},
  \bibinfo {author} {\bibfnamefont {Y.-f.}\ \bibnamefont {He}}, \bibinfo
  {author} {\bibfnamefont {F.-g.}\ \bibnamefont {Li}}, \ and\ \bibinfo {author}
  {\bibfnamefont {W.-r.}\ \bibnamefont {Zhong}},\ }\href
  {https://doi.org/10.1103/PhysRevE.88.062129} {\bibfield  {journal} {\bibinfo
  {journal} {Phys. Rev. E}\ }\textbf {\bibinfo {volume} {88}},\ \bibinfo
  {pages} {062129} (\bibinfo {year} {2013})}\BibitemShut {NoStop}%
\bibitem [{\citenamefont {Fily}\ \emph {et~al.}(2014)\citenamefont {Fily},
  \citenamefont {Baskaran},\ and\ \citenamefont {Hagan}}]{fily2014dynamics}%
  \BibitemOpen
  \bibfield  {author} {\bibinfo {author} {\bibfnamefont {Y.}~\bibnamefont
  {Fily}}, \bibinfo {author} {\bibfnamefont {A.}~\bibnamefont {Baskaran}}, \
  and\ \bibinfo {author} {\bibfnamefont {M.~F.}\ \bibnamefont {Hagan}},\ }\href
  {https://doi.org/10.1039/C4SM00975D} {\bibfield  {journal} {\bibinfo
  {journal} {Soft Matter}\ }\textbf {\bibinfo {volume} {10}},\ \bibinfo {pages}
  {5609} (\bibinfo {year} {2014})}\BibitemShut {NoStop}%
\bibitem [{\citenamefont {Basu}\ \emph {et~al.}(2018)\citenamefont {Basu},
  \citenamefont {Majumdar}, \citenamefont {Rosso},\ and\ \citenamefont
  {Schehr}}]{basu2018}%
  \BibitemOpen
  \bibfield  {author} {\bibinfo {author} {\bibfnamefont {U.}~\bibnamefont
  {Basu}}, \bibinfo {author} {\bibfnamefont {S.~N.}\ \bibnamefont {Majumdar}},
  \bibinfo {author} {\bibfnamefont {A.}~\bibnamefont {Rosso}}, \ and\ \bibinfo
  {author} {\bibfnamefont {G.}~\bibnamefont {Schehr}},\ }\href
  {https://link.aps.org/doi/10.1103/PhysRevE.98.062121} {\bibfield  {journal}
  {\bibinfo  {journal} {Phys. Rev. E}\ }\textbf {\bibinfo {volume} {98}},\
  \bibinfo {pages} {062121} (\bibinfo {year} {2018})}\BibitemShut {NoStop}%
\bibitem [{\citenamefont {Dauchot}\ and\ \citenamefont
  {D\'emery}(2019)}]{dauchot2019}%
  \BibitemOpen
  \bibfield  {author} {\bibinfo {author} {\bibfnamefont {O.}~\bibnamefont
  {Dauchot}}\ and\ \bibinfo {author} {\bibfnamefont {V.}~\bibnamefont
  {D\'emery}},\ }\href
  {https://link.aps.org/doi/10.1103/PhysRevLett.122.068002} {\bibfield
  {journal} {\bibinfo  {journal} {Phys. Rev. Lett.}\ }\textbf {\bibinfo
  {volume} {122}},\ \bibinfo {pages} {068002} (\bibinfo {year}
  {2019})}\BibitemShut {NoStop}%
\bibitem [{\citenamefont {Kranz}\ and\ \citenamefont
  {Golestanian}(2019)}]{kranz2019}%
  \BibitemOpen
  \bibfield  {author} {\bibinfo {author} {\bibfnamefont {W.~T.}\ \bibnamefont
  {Kranz}}\ and\ \bibinfo {author} {\bibfnamefont {R.}~\bibnamefont
  {Golestanian}},\ }\href {https://doi.org/10.1063/1.5081122} {\bibfield
  {journal} {\bibinfo  {journal} {J. Chem. Phys.}\ }\textbf {\bibinfo {volume}
  {150}},\ \bibinfo {pages} {214111} (\bibinfo {year} {2019})},\ \Eprint
  {http://arxiv.org/abs/https://doi.org/10.1063/1.5081122}
  {https://doi.org/10.1063/1.5081122} \BibitemShut {NoStop}%
\bibitem [{\citenamefont {Liverpool}(2003)}]{liverpool2003}%
  \BibitemOpen
  \bibfield  {author} {\bibinfo {author} {\bibfnamefont {T.~B.}\ \bibnamefont
  {Liverpool}},\ }\href {https://doi.org/10.1103/physreve.67.031909} {\bibfield
   {journal} {\bibinfo  {journal} {Phys. Rev. E}\ }\textbf {\bibinfo {volume}
  {67}} (\bibinfo {year} {2003})}\BibitemShut {NoStop}%
\bibitem [{\citenamefont {Gao}\ \emph {et~al.}(2017)\citenamefont {Gao},
  \citenamefont {Betterton}, \citenamefont {Jhang},\ and\ \citenamefont
  {Shelley}}]{gao2017}%
  \BibitemOpen
  \bibfield  {author} {\bibinfo {author} {\bibfnamefont {T.}~\bibnamefont
  {Gao}}, \bibinfo {author} {\bibfnamefont {M.~D.}\ \bibnamefont {Betterton}},
  \bibinfo {author} {\bibfnamefont {A.-S.}\ \bibnamefont {Jhang}}, \ and\
  \bibinfo {author} {\bibfnamefont {M.~J.}\ \bibnamefont {Shelley}},\ }\href
  {https://link.aps.org/doi/10.1103/PhysRevFluids.2.093302} {\bibfield
  {journal} {\bibinfo  {journal} {Phys. Rev. Fluids}\ }\textbf {\bibinfo
  {volume} {2}},\ \bibinfo {pages} {093302} (\bibinfo {year}
  {2017})}\BibitemShut {NoStop}%
\bibitem [{\citenamefont {Eisenstecken}\ \emph {et~al.}(2016)\citenamefont
  {Eisenstecken}, \citenamefont {Gompper},\ and\ \citenamefont
  {Winkler}}]{eisenstecken2016}%
  \BibitemOpen
  \bibfield  {author} {\bibinfo {author} {\bibfnamefont {T.}~\bibnamefont
  {Eisenstecken}}, \bibinfo {author} {\bibfnamefont {G.}~\bibnamefont
  {Gompper}}, \ and\ \bibinfo {author} {\bibfnamefont {R.}~\bibnamefont
  {Winkler}},\ }\href {http://www.mdpi.com/2073-4360/8/8/304} {\bibfield
  {journal} {\bibinfo  {journal} {Polymers}\ }\textbf {\bibinfo {volume} {8}},\
  \bibinfo {pages} {304} (\bibinfo {year} {2016})}\BibitemShut {NoStop}%
\bibitem [{\citenamefont {Eisenstecken}\ \emph {et~al.}(2017)\citenamefont
  {Eisenstecken}, \citenamefont {Gompper},\ and\ \citenamefont
  {Winkler}}]{eisenstecken2017}%
  \BibitemOpen
  \bibfield  {author} {\bibinfo {author} {\bibfnamefont {T.}~\bibnamefont
  {Eisenstecken}}, \bibinfo {author} {\bibfnamefont {G.}~\bibnamefont
  {Gompper}}, \ and\ \bibinfo {author} {\bibfnamefont {R.~G.}\ \bibnamefont
  {Winkler}},\ }\href {http://aip.scitation.org/doi/10.1063/1.4981012}
  {\bibfield  {journal} {\bibinfo  {journal} {J. Chem. Phys.}\ }\textbf
  {\bibinfo {volume} {146}},\ \bibinfo {pages} {154903} (\bibinfo {year}
  {2017})}\BibitemShut {NoStop}%
\bibitem [{\citenamefont {Osmanović}\ and\ \citenamefont
  {Rabin}(2017)}]{osmanovic2017}%
  \BibitemOpen
  \bibfield  {author} {\bibinfo {author} {\bibfnamefont {D.}~\bibnamefont
  {Osmanović}}\ and\ \bibinfo {author} {\bibfnamefont {Y.}~\bibnamefont
  {Rabin}},\ }\href {http://xlink.rsc.org/?DOI=C6SM02722A} {\bibfield
  {journal} {\bibinfo  {journal} {Soft Matter}\ }\textbf {\bibinfo {volume}
  {13}},\ \bibinfo {pages} {963} (\bibinfo {year} {2017})}\BibitemShut
  {NoStop}%
\bibitem [{\citenamefont {Osmanović}(2018)}]{osmanovic2018}%
  \BibitemOpen
  \bibfield  {author} {\bibinfo {author} {\bibfnamefont {D.}~\bibnamefont
  {Osmanović}},\ }\href {https://doi.org/10.1063/1.5045686} {\bibfield
  {journal} {\bibinfo  {journal} {J. Chem. Phys.}\ }\textbf {\bibinfo {volume}
  {149}},\ \bibinfo {pages} {164911} (\bibinfo {year} {2018})}\BibitemShut
  {NoStop}%
\bibitem [{\citenamefont {De~Canio}\ \emph {et~al.}(2017)\citenamefont
  {De~Canio}, \citenamefont {Lauga},\ and\ \citenamefont
  {Goldstein}}]{de_canio2017}%
  \BibitemOpen
  \bibfield  {author} {\bibinfo {author} {\bibfnamefont {G.}~\bibnamefont
  {De~Canio}}, \bibinfo {author} {\bibfnamefont {E.}~\bibnamefont {Lauga}}, \
  and\ \bibinfo {author} {\bibfnamefont {R.~E.}\ \bibnamefont {Goldstein}},\
  }\href {https://doi.org/10.1098/rsif.2017.0491} {\bibfield  {journal}
  {\bibinfo  {journal} {J. R. Soc., Interface}\ }\textbf {\bibinfo {volume}
  {14}},\ \bibinfo {pages} {20170491} (\bibinfo {year} {2017})}\BibitemShut
  {NoStop}%
\bibitem [{\citenamefont {Weber}\ \emph {et~al.}(2010)\citenamefont {Weber},
  \citenamefont {Theriot},\ and\ \citenamefont {Spakowitz}}]{weber2010}%
  \BibitemOpen
  \bibfield  {author} {\bibinfo {author} {\bibfnamefont {S.~C.}\ \bibnamefont
  {Weber}}, \bibinfo {author} {\bibfnamefont {J.~A.}\ \bibnamefont {Theriot}},
  \ and\ \bibinfo {author} {\bibfnamefont {A.~J.}\ \bibnamefont {Spakowitz}},\
  }\href {https://link.aps.org/doi/10.1103/PhysRevE.82.011913} {\bibfield
  {journal} {\bibinfo  {journal} {Phys. Rev. E}\ }\textbf {\bibinfo {volume}
  {82}},\ \bibinfo {pages} {011913} (\bibinfo {year} {2010})}\BibitemShut
  {NoStop}%
\bibitem [{\citenamefont {Ghosh}\ and\ \citenamefont {Gov}(2014)}]{ghosh2014}%
  \BibitemOpen
  \bibfield  {author} {\bibinfo {author} {\bibfnamefont {A.}~\bibnamefont
  {Ghosh}}\ and\ \bibinfo {author} {\bibfnamefont {N.}~\bibnamefont {Gov}},\
  }\href {https://linkinghub.elsevier.com/retrieve/pii/S0006349514007796}
  {\bibfield  {journal} {\bibinfo  {journal} {Biophys. J.}\ }\textbf {\bibinfo
  {volume} {107}},\ \bibinfo {pages} {1065} (\bibinfo {year}
  {2014})}\BibitemShut {NoStop}%
\bibitem [{\citenamefont {Di~Pierro}\ \emph {et~al.}(2018)\citenamefont
  {Di~Pierro}, \citenamefont {Potoyan}, \citenamefont {Wolynes},\ and\
  \citenamefont {Onuchic}}]{di_pierro2018}%
  \BibitemOpen
  \bibfield  {author} {\bibinfo {author} {\bibfnamefont {M.}~\bibnamefont
  {Di~Pierro}}, \bibinfo {author} {\bibfnamefont {D.~A.}\ \bibnamefont
  {Potoyan}}, \bibinfo {author} {\bibfnamefont {P.~G.}\ \bibnamefont
  {Wolynes}}, \ and\ \bibinfo {author} {\bibfnamefont {J.~N.}\ \bibnamefont
  {Onuchic}},\ }\href {https://www.pnas.org/content/115/30/7753} {\bibfield
  {journal} {\bibinfo  {journal} {Proc. Natl. Acad. Sci.}\ }\textbf {\bibinfo
  {volume} {115}},\ \bibinfo {pages} {7753} (\bibinfo {year}
  {2018})}\BibitemShut {NoStop}%
\bibitem [{\citenamefont {Bianco}\ \emph {et~al.}(2018)\citenamefont {Bianco},
  \citenamefont {Locatelli},\ and\ \citenamefont {Malgaretti}}]{bianco2018}%
  \BibitemOpen
  \bibfield  {author} {\bibinfo {author} {\bibfnamefont {V.}~\bibnamefont
  {Bianco}}, \bibinfo {author} {\bibfnamefont {E.}~\bibnamefont {Locatelli}}, \
  and\ \bibinfo {author} {\bibfnamefont {P.}~\bibnamefont {Malgaretti}},\
  }\href {https://link.aps.org/doi/10.1103/PhysRevLett.121.217802} {\bibfield
  {journal} {\bibinfo  {journal} {Phys. Rev. Lett.}\ }\textbf {\bibinfo
  {volume} {121}},\ \bibinfo {pages} {217802} (\bibinfo {year}
  {2018})}\BibitemShut {NoStop}%
\bibitem [{\citenamefont {Chaki}\ and\ \citenamefont
  {Chakrabarti}(2019)}]{chaki2019}%
  \BibitemOpen
  \bibfield  {author} {\bibinfo {author} {\bibfnamefont {S.}~\bibnamefont
  {Chaki}}\ and\ \bibinfo {author} {\bibfnamefont {R.}~\bibnamefont
  {Chakrabarti}},\ }\href {https://doi.org/10.1063/1.5086152} {\bibfield
  {journal} {\bibinfo  {journal} {J. Chem. Phys.}\ }\textbf {\bibinfo {volume}
  {150}},\ \bibinfo {pages} {094902} (\bibinfo {year} {2019})}\BibitemShut
  {NoStop}%
\bibitem [{\citenamefont {Anand}\ and\ \citenamefont
  {Singh}(2018)}]{anand2018}%
  \BibitemOpen
  \bibfield  {author} {\bibinfo {author} {\bibfnamefont {S.~K.}\ \bibnamefont
  {Anand}}\ and\ \bibinfo {author} {\bibfnamefont {S.~P.}\ \bibnamefont
  {Singh}},\ }\href {https://link.aps.org/doi/10.1103/PhysRevE.98.042501}
  {\bibfield  {journal} {\bibinfo  {journal} {Phys. Rev. E}\ }\textbf {\bibinfo
  {volume} {98}},\ \bibinfo {pages} {042501} (\bibinfo {year}
  {2018})}\BibitemShut {NoStop}%
\bibitem [{\citenamefont {Das}\ and\ \citenamefont {Cacciuto}(2019)}]{das2019}%
  \BibitemOpen
  \bibfield  {author} {\bibinfo {author} {\bibfnamefont {S.}~\bibnamefont
  {Das}}\ and\ \bibinfo {author} {\bibfnamefont {A.}~\bibnamefont {Cacciuto}},\
  }\href {https://doi.org/10.1103/physrevlett.123.087802} {\bibfield  {journal}
  {\bibinfo  {journal} {Phys. Rev. Lett.}\ }\textbf {\bibinfo {volume} {123}}
  (\bibinfo {year} {2019})}\BibitemShut {NoStop}%
\bibitem [{\citenamefont {Harder}\ \emph {et~al.}(2014)\citenamefont {Harder},
  \citenamefont {Valeriani},\ and\ \citenamefont {Cacciuto}}]{harder2014}%
  \BibitemOpen
  \bibfield  {author} {\bibinfo {author} {\bibfnamefont {J.}~\bibnamefont
  {Harder}}, \bibinfo {author} {\bibfnamefont {C.}~\bibnamefont {Valeriani}}, \
  and\ \bibinfo {author} {\bibfnamefont {A.}~\bibnamefont {Cacciuto}},\ }\href
  {https://link.aps.org/doi/10.1103/PhysRevE.90.062312} {\bibfield  {journal}
  {\bibinfo  {journal} {Phys. Rev. E}\ }\textbf {\bibinfo {volume} {90}},\
  \bibinfo {pages} {062312} (\bibinfo {year} {2014})}\BibitemShut {NoStop}%
\bibitem [{\citenamefont {{Doi}}\ and\ \citenamefont
  {{Edwards}}(2007)}]{doi2007}%
  \BibitemOpen
  \bibfield  {author} {\bibinfo {author} {\bibfnamefont {M.}~\bibnamefont
  {{Doi}}}\ and\ \bibinfo {author} {\bibfnamefont {S.~F.}\ \bibnamefont
  {{Edwards}}},\ }\href@noop {} {\emph {\bibinfo {title} {The Theory of Polymer
  Dynamics}}},\ Vol.~\bibinfo {volume} {73}\ (\bibinfo  {publisher} {Oxford
  University Press},\ \bibinfo {year} {2007})\BibitemShut {NoStop}%
\bibitem [{\citenamefont {Pringle}\ \emph {et~al.}(2013)\citenamefont
  {Pringle}, \citenamefont {Muthukumar}, \citenamefont {Tan}, \citenamefont
  {Crankshaw}, \citenamefont {Conway},\ and\ \citenamefont
  {Ross}}]{pringle2013}%
  \BibitemOpen
  \bibfield  {author} {\bibinfo {author} {\bibfnamefont {J.}~\bibnamefont
  {Pringle}}, \bibinfo {author} {\bibfnamefont {A.}~\bibnamefont {Muthukumar}},
  \bibinfo {author} {\bibfnamefont {A.}~\bibnamefont {Tan}}, \bibinfo {author}
  {\bibfnamefont {L.}~\bibnamefont {Crankshaw}}, \bibinfo {author}
  {\bibfnamefont {L.}~\bibnamefont {Conway}}, \ and\ \bibinfo {author}
  {\bibfnamefont {J.~L.}\ \bibnamefont {Ross}},\ }\href
  {https://doi.org/10.1088/0953-8984/25/37/374103} {\bibfield  {journal}
  {\bibinfo  {journal} {J. Phys. Condens. Matter}\ }\textbf {\bibinfo {volume}
  {25}},\ \bibinfo {pages} {374103} (\bibinfo {year} {2013})}\BibitemShut
  {NoStop}%
\bibitem [{\citenamefont {Farhadi}\ \emph {et~al.}(2018)\citenamefont
  {Farhadi}, \citenamefont {Rosario}, \citenamefont {Debold}, \citenamefont
  {Baskaran},\ and\ \citenamefont {Ross}}]{farhadi2018}%
  \BibitemOpen
  \bibfield  {author} {\bibinfo {author} {\bibfnamefont {L.}~\bibnamefont
  {Farhadi}}, \bibinfo {author} {\bibfnamefont {C.~F.~D.}\ \bibnamefont
  {Rosario}}, \bibinfo {author} {\bibfnamefont {E.~P.}\ \bibnamefont {Debold}},
  \bibinfo {author} {\bibfnamefont {A.}~\bibnamefont {Baskaran}}, \ and\
  \bibinfo {author} {\bibfnamefont {J.~L.}\ \bibnamefont {Ross}},\ }\href
  {https://doi.org/10.3389/fphy.2018.00075} {\bibfield  {journal} {\bibinfo
  {journal} {Front. Phys.}\ }\textbf {\bibinfo {volume} {6}} (\bibinfo {year}
  {2018})}\BibitemShut {NoStop}%
\bibitem [{\citenamefont {Sumino}\ \emph {et~al.}(2012)\citenamefont {Sumino},
  \citenamefont {Nagai}, \citenamefont {Shitaka}, \citenamefont {Tanaka},
  \citenamefont {Yoshikawa}, \citenamefont {Chat{\'{e}}},\ and\ \citenamefont
  {Oiwa}}]{sumino2012}%
  \BibitemOpen
  \bibfield  {author} {\bibinfo {author} {\bibfnamefont {Y.}~\bibnamefont
  {Sumino}}, \bibinfo {author} {\bibfnamefont {K.~H.}\ \bibnamefont {Nagai}},
  \bibinfo {author} {\bibfnamefont {Y.}~\bibnamefont {Shitaka}}, \bibinfo
  {author} {\bibfnamefont {D.}~\bibnamefont {Tanaka}}, \bibinfo {author}
  {\bibfnamefont {K.}~\bibnamefont {Yoshikawa}}, \bibinfo {author}
  {\bibfnamefont {H.}~\bibnamefont {Chat{\'{e}}}}, \ and\ \bibinfo {author}
  {\bibfnamefont {K.}~\bibnamefont {Oiwa}},\ }\href
  {https://doi.org/10.1038/nature10874} {\bibfield  {journal} {\bibinfo
  {journal} {Nature}\ }\textbf {\bibinfo {volume} {483}},\ \bibinfo {pages}
  {448} (\bibinfo {year} {2012})}\BibitemShut {NoStop}%
\bibitem [{\citenamefont {Schaller}\ \emph {et~al.}(2010)\citenamefont
  {Schaller}, \citenamefont {Weber}, \citenamefont {Semmrich}, \citenamefont
  {Frey},\ and\ \citenamefont {Bausch}}]{schaller2010}%
  \BibitemOpen
  \bibfield  {author} {\bibinfo {author} {\bibfnamefont {V.}~\bibnamefont
  {Schaller}}, \bibinfo {author} {\bibfnamefont {C.}~\bibnamefont {Weber}},
  \bibinfo {author} {\bibfnamefont {C.}~\bibnamefont {Semmrich}}, \bibinfo
  {author} {\bibfnamefont {E.}~\bibnamefont {Frey}}, \ and\ \bibinfo {author}
  {\bibfnamefont {A.~R.}\ \bibnamefont {Bausch}},\ }\href
  {https://doi.org/10.1038/nature09312} {\bibfield  {journal} {\bibinfo
  {journal} {Nature}\ }\textbf {\bibinfo {volume} {467}},\ \bibinfo {pages}
  {73} (\bibinfo {year} {2010})}\BibitemShut {NoStop}%
\bibitem [{\citenamefont {Ghosh}\ \emph {et~al.}(2015)\citenamefont {Ghosh},
  \citenamefont {Li}, \citenamefont {Marchegiani},\ and\ \citenamefont
  {Marchesoni}}]{ghosh2015}%
  \BibitemOpen
  \bibfield  {author} {\bibinfo {author} {\bibfnamefont {P.~K.}\ \bibnamefont
  {Ghosh}}, \bibinfo {author} {\bibfnamefont {Y.}~\bibnamefont {Li}}, \bibinfo
  {author} {\bibfnamefont {G.}~\bibnamefont {Marchegiani}}, \ and\ \bibinfo
  {author} {\bibfnamefont {F.}~\bibnamefont {Marchesoni}},\ }\href
  {http://aip.scitation.org/doi/10.1063/1.4936624} {\bibfield  {journal}
  {\bibinfo  {journal} {J. Chem. Phys.}\ }\textbf {\bibinfo {volume} {143}},\
  \bibinfo {pages} {211101} (\bibinfo {year} {2015})}\BibitemShut {NoStop}%
\bibitem [{\citenamefont {Weeks}\ \emph {et~al.}(1971)\citenamefont {Weeks},
  \citenamefont {Chandler},\ and\ \citenamefont {Andersen}}]{weeks1971}%
  \BibitemOpen
  \bibfield  {author} {\bibinfo {author} {\bibfnamefont {J.~D.}\ \bibnamefont
  {Weeks}}, \bibinfo {author} {\bibfnamefont {D.}~\bibnamefont {Chandler}}, \
  and\ \bibinfo {author} {\bibfnamefont {H.~C.}\ \bibnamefont {Andersen}},\
  }\href {https://doi.org/10.1063/1.1674820} {\bibfield  {journal} {\bibinfo
  {journal} {J. Chem. Phys.}\ }\textbf {\bibinfo {volume} {54}},\ \bibinfo
  {pages} {5237} (\bibinfo {year} {1971})}\BibitemShut {NoStop}%
\end{thebibliography}%


\appendix

\section{Mean Square Bond Length with Active Beads}
\label{app:MeanSquareBondLength}

If we imagine a Rouse chain whose beads are the active components, then we must
choose a canonical tangent direction for the bead to exert forces in, which must
depend on the bond vectors associated to that bead. In particular, we can
parameterize the possible tangent vectors $\vb{t}_n$ by a parameter $\nu$:
\begin{equation}
    \vb{t}_n(\nu) = \frac{\nu + 1}{2} (\vb{r}_{n+1} - \vb{r}_n)
    - \frac{\nu - 1}{2} (\vb{r}_{n} - \vb{r}_{n-1}).
\end{equation}

Consider a Rouse filament consisting of only two beads (and therefore only one
bond vector $\vb{b} = \vb{r}_2 - \vb{r}_1$). The equations of motion for these
beads takes the simple form
\begin{eqnarray*}
    \gamma \partial_t \vb{r}_1 &=& k \vb{b} + \frac{1}{2} \fa (\nu + 1) \vb{b} +
    \vb{\xi}_1 \\
    \gamma \partial_t \vb{r}_2 &=& -k\vb{b} - \frac{1}{2} \fa (\nu - 1) \vb{b} +
    \vb{\xi}_2.
\end{eqnarray*}
Subtracting the latter from the former, we obtain an equation of motion for the
bond vector:
\begin{equation}
    \gamma \partial_t \vb{b} = -2(k/\gamma)(1 + \alpha \nu / N) \vb{b}
    + \vb{\zeta},
\end{equation}
where we have introduced $\alpha = fN/2k$ and $\vb{\zeta} = \vb{\xi}_2 -
\vb{\xi}_1$. This can be readily solved to find
\begin{equation}
    \vb{b}(t)
    = \vb{b}(0) e^{-t/\tau} + \int_0^t \dd{s} e^{-(t-s)/\tau} \zeta(s),
\end{equation}
where
\begin{equation}
    \tau = \frac{\gamma}{2k(1 + \alpha \nu / N)}.
\end{equation}
Squaring and averaging, we find that in the long time limit
\begin{equation}
    b^2
    = \lim_{t \to \infty} \ev{b(t)^2}
    = \frac{2 d \kt \tau}{\gamma}
    = \frac{b_0^2}{1 + \alpha \nu / N},
\end{equation}
where $b_0^2 = d \kt / k$ is the mean square bond length of the passive Rouse
chain. Thus, we find that activity can lead to compression ($\nu < 0$) or
expansion ($\nu > 0$) of the bonds. The bond length is unchanged if $\nu = 0$;
in fact, $\nu = 0$ is equivalent to the case where the bonds are the active
agents, as used in the main text.

\begin{figure}
    \centering
    \includegraphics{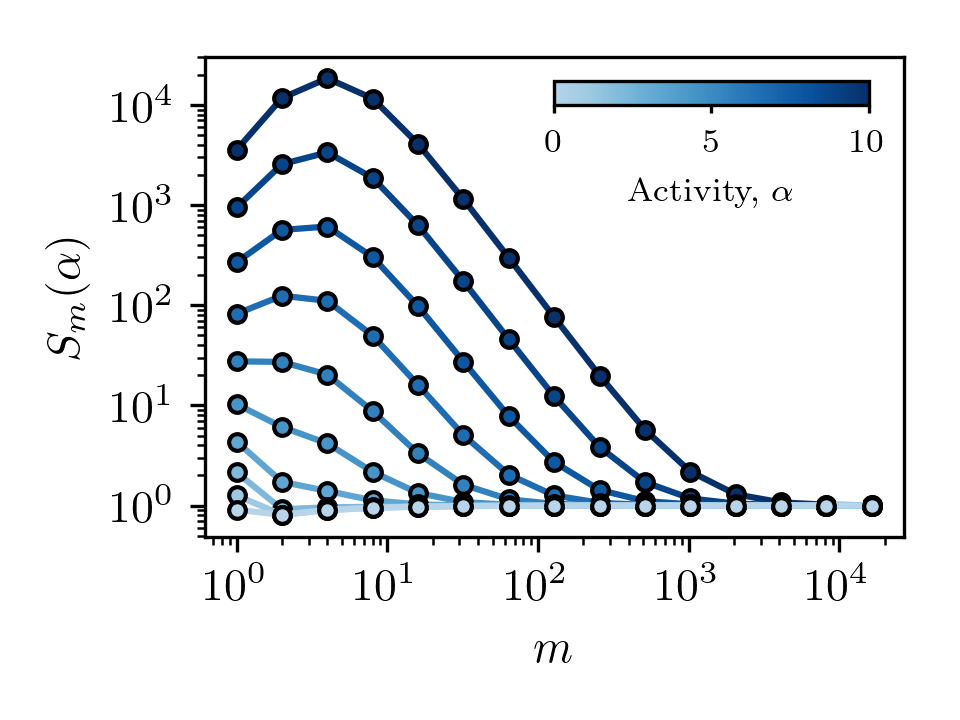}
    \caption{Numerical evaluation of Eq.~\eqref{eq:EndToEndPartialSum}
    for various values of the activity parameter $\alpha$ as a function of the
    number of terms in the sum, $m$. The sum converges to 1 for all tested
    activities, for sufficiently large $m$, with more terms required for larger
    $\alpha$ values.}
    \label{fig:EndToEndPartialSum}
\end{figure}

\section{Mean Square Displacement and the Diffusion Coefficient}
\label{app:MeanSquareDisplacement}

As discussed in the main text, the MSD can be written as
\begin{eqnarray*}
    \text{MSD} &=&
    \sum_{p,q} \bar{\phi}_p \bar{\phi}_q \ev{\vb{c}_p(t) \cdot \vb{c}_q(t)}.
\end{eqnarray*}
Using the definition of $\vb{c}_p(t)$ given in Eq.~\eqref{eq:CpSolution}, we
have that
\begin{align}
    \langle &\vb{c}_p(t + \tau) \cdot \vb{c}_q(t) \rangle \nonumber \\
    &= dG_{pq} \times
    \begin{cases}
        t & p = q = 0 \\
        e^{-\lambda_p^2 \tau}
        \dfrac{1 - e^{-(\lambda_p^2 + \lambda_q^2)t}}
        {\lambda_p^2 + \lambda_q^2} & \text{otherwise}
    \end{cases}
\end{align}
Thus, for $\tau = 0$, this correlation function approaches a constant if either
$p > 0$ or $q > 0$. The only term that grows without bound is the $p = q = 0$
term. The other terms can be collected into the function $F(t)$ as used in
Eq.~\eqref{eq:MeanSquareDisplacement}. In particular, we have that
\begin{equation}
    \lim_{t \to \infty} \frac{\ev{\vb{c}_p(t) \cdot \vb{c}_q(t)}}{t}
    = \delta_{p,0} \delta_{q,0}
\end{equation}
which we use to compute the diffusion coefficient as in
Eq.~\eqref{eq:DiffusionCoefficient}.

\begin{figure}
    \centering
    \includegraphics{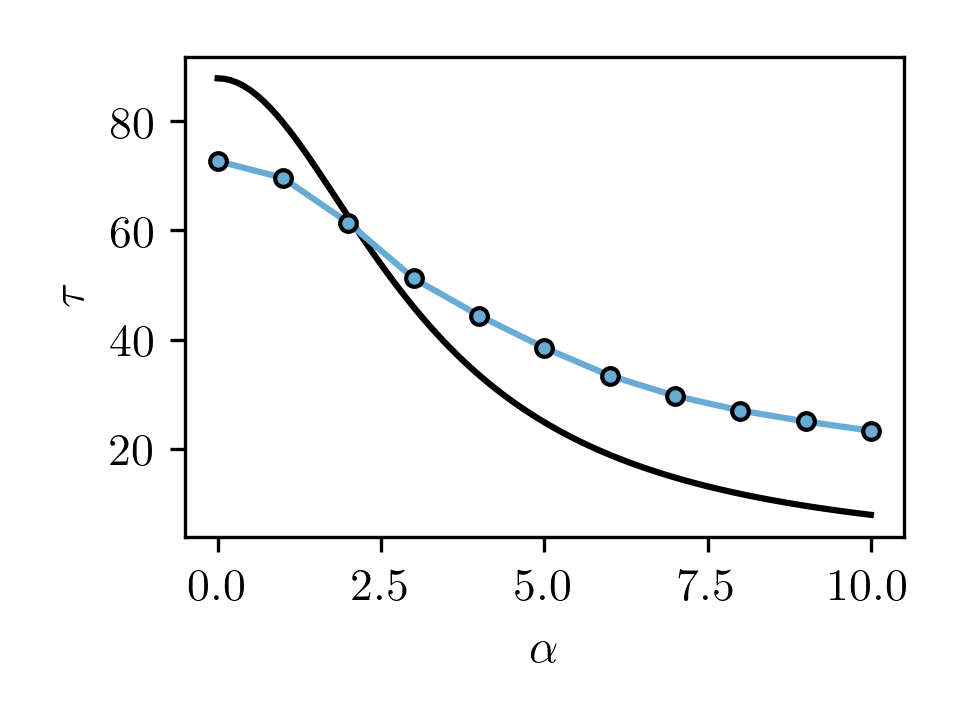}
    \caption{Measured relaxation time (blue dots) compared to that predicted
        from Eq.~\eqref{eq:RelaxationTime} (black line). As expected, activity
        reduces the relaxation time; however, we only observe qualitative
        agreement between simulation and theory.}
    \label{fig:RotationalRelaxationTime}
\end{figure}

\section{Effect of Inertia}
\label{app:EffectOfInertia}

The equation of motion for an active Rouse chain with inertia (in the continuum
limit) would be
\[
    m\pdv[2]{\vb{r}}{t} + \gamma \pdv{\vb{r}}{t}
    = k \pdv[2]{\vb{r}}{n} + f_a \pdv{\vb{r}}{n} 
    + \sqrt{2 \gamma k_B T} \vb{\xi}(n, t).
\]
We can non-dimensionalize this in the same manner as in the main text, measuring
time in units of $d \gamma / k$, distance in units of $b_0 = \sqrt{dk_B T/k}$,
and energy in units of $k_B T$. Defining $\alpha = f_a N b_0^2 / 2 d k_B T$, we
find
\[
    \tau \pdv[2]{\vb{r}}{t} + \pdv{\vb{r}}{t}
    = d \pdv[2]{\vb{r}}{n} + \frac{2d\alpha}{N} \pdv{\vb{r}}{n} 
    + \sqrt{2} \vb{\xi}(n, t),
\]
where $\tau = m k / d \gamma^2$ is an inertial timescale. As before, we can
write the solutions in terms of the eigenfunctions $\phi_p(n)$ as
\[
    \vb{r}(n, t) = \sum_{p=0}^{\infty} \vb{c}_p(t) \phi_p(n).
\]
The coefficients $\vb{c}_p(t)$ now takes the form
\[
    \vb{c}_p(t) = \int_0^t \dd{s} H_p(t - s) \vb{\xi}_p(s),
\]
where
\[
    H_p(t) = e^{-t/2\tau} \times \frac{2\sinh(z_p t / 2 \tau)}{z_p},
\]
$z_p = \sqrt{1 - 4 \lambda_p^2 \tau}$, and $\lambda_p^2 = (d/N^2)[\pi^2 p^2 + (1
- \delta_{p,0})\alpha^2]$, as in the main text. Note that for $4 \lambda_p^2
\tau > 1$, $z_p$ becomes imaginary and oscillatory behavior sets in. This is in
contrast to the results found in the overdamped limit, where no oscillatory
motion is present.

We can compute
\begin{align}
    \lim_{t \to \infty} &\ev{\vb{c}_p(t+\Delta) \cdot \vb{c}_q(t)} \nonumber \\
    &= dG_{pq} \times \begin{cases}
        t & p = q = 0 \\
        F_{pq}(\Delta) & \text{otherwise}
    \end{cases}
\end{align}
where
\begin{widetext}
\begin{equation}
    F_{pq}(\Delta)
    = 8 \tau e^{-\Delta/2\tau} \left\{
        \frac{(z_p^2 - z_q^2 + 4) \sinh(z_p \Delta/2\tau) 
            + 4 z_p \cosh(z_p \Delta / 2\tau)}
        {z_p (4 - (z_p - z_q)^2)(4 - (z_p + z_q)^2)}
    \right\}
\end{equation}
\end{widetext}
We therefore see that, in the long time limit, the $p = q = 0$ term is again the
only term that contributes to the MSD, and so the addition of inertia does not
affect the diffusive behavior of an active Rouse chain.

We can determine the effect of the active driving at short times by computing
\begin{align}
    \text{MSD}(\Delta)
    &= \lim_{t \to \infty} \ev{(\vb{X}(t + \Delta) - \vb{X}(t))^2} \nonumber \\
    &= d G_{00} \phi_0^2 \Delta 
    - \sideset{}{'}\sum_{p,q} [f_{pq}(\Delta) + f_{qp}(\Delta)],
\end{align}
where $f_{pq}(\Delta) = F_{pq}(\Delta) - F_{pq}(0)$, and primed sums indicate
summing over all terms except the $p = q = 0$ term. To second order, we have
\[
    f_{pq}(\Delta) \approx
    \frac{2}{\tau} \times
    \frac{(2 \Delta \tau + \Delta^2)(z_p^2 - z_q^2) - 2 \Delta^2}
    {(4 - (z_p - z_q)^2)(4 - (z_p + z_q)^2)}.
\]
Noting the terms that are asymmetric in $p$ and $q$, we can write
\[
    f_{pq}(\Delta) + f_{qp}(\Delta)
    =  - \Delta^2 b_{pq}
\]
where
\begin{align}
    b_{pq} &= \frac{8/\tau}{(4 - (z_p - z_q)^2)(4 - (z_p + z_q)^2)} \nonumber \\
    &= \frac{1}{2\tau} \times 
    \frac{1}
    {\tau^2 (\lambda_p^2 - \lambda_q^2)^2 + 2 \tau (\lambda_p^2 + \lambda_q^2)},
    \label{eq:BallisticTerm}
\end{align}
where in the second line we have used the definition of $z_p$ to rewrite the
expression in terms of $\lambda_p$. Defining
\begin{equation} \label{eq:BallisticCoefficient}
    B(\alpha) = \sideset{}{'}\sum_{p,q} b_{pq},
\end{equation}
we can write the short-time MSD as
\begin{equation}
    \text{MSD}(\Delta) \approx 2d D(\alpha) \Delta + B(\alpha) \Delta^2,
\end{equation}
where $D(\alpha)$ is the diffusion coefficient from
Eq.~\eqref{eq:DiffusionCoefficient} and $B(\alpha)$ is an activity-dependent
coefficient describing the ballistic motion of the system at short times. It's
not clear if the sum in Eq.~\eqref{eq:BallisticCoefficient} can be computed
analytically; however, from the structure of the terms in
Eq.~\eqref{eq:BallisticTerm} we can see that increasing activity
\textit{reduces} the value of $B(\alpha)$.

\section{Radius of Gyration}
\label{app:RadiusOfGyration}

In addition to the mean-squared end-to-end distance, $\ev{L^2}$, the radius of
gyration, $\Rg$, is a common descriptor of polymer conformations measured in
experiments. It is calculated as
\begin{eqnarray}
    \Rg^2 &=& \frac{1}{N} \int_0^N \dd{n} (\vb{r}(n,t) - \vb{X}(t))^2 
    \\
    &=& d \sum_{p,q>0} \frac{G_{pq}}{\lambda_p^2 + \lambda_q^2} \left[
        \frac{1}{N} \int_0^N \phi_p(n) \phi_q(n) - \bar{\phi}_p \bar{\phi}_q    
    \right] \nonumber.
\end{eqnarray}
As with Eq.~\eqref{eq:EndToEndDistanceExpression}, this sum is not amenable to
analytic computation, and must be numerically summed. We find that $\Rg$ is also
independent of activity, and converges slowly to
\begin{equation}
    \Rg^2 \approx N/12,
\end{equation}
which is the same as that of the passive Rouse chain.

\section{Mean Square End-to-End Distance Computation}
\label{app:MeanSquareEndToEndDistance}

We claim that Eq.~\eqref{eq:EndToEndDistanceExpression} converges such that
Eq.~\eqref{eq:EndToEndDistance} holds. Here, we give numerical evidence of our
claim. Note that, after evaluating Eq.~\eqref{eq:NoiseModeCorrelation}, we can
explicitly write the sum as
\begin{equation}
    \ev{L^2}/N = \sum_{p,q > 0} s_{p,q},
\end{equation}
where
\begin{widetext}
\begin{equation}
    s_{p,q} = 16\pi^4 \alpha \times
    \frac
    {
        p^2 q^2 [(-1)^p e^{-\alpha}-1]
        [(-1)^q e^{-\alpha} - 1]
        [(-1)^{p+q} e^{2\alpha} - 1]
    }
    {
        (\pi^2 p^2 + \alpha^2)
        (\pi^2 q^2 + \alpha^2)
        (\pi^2 (p + q)^2 + 4\alpha^2)
        (\pi^2 (p - q)^2 + 4\alpha^2)
    }.
\end{equation}
\end{widetext}
We define the partial sum $S_m(\alpha)$ as
\begin{equation}
    \label{eq:EndToEndPartialSum}
    S_m(\alpha) = \sum_{p=1}^m \sum_{q=1}^m s_{p,q}
\end{equation}
so that
\begin{equation}
    \ev{L^2}/N = \lim_{m \to \infty} S_m(\alpha).
\end{equation}
In Fig.~\ref{fig:EndToEndPartialSum}, we plot $S_m(\alpha)$ against $m$ for a
few values of $\alpha$. In all cases, we see that $S_m(\alpha) \to 1$ as $m$
increases, though for $\alpha > 1$ the partial sum can grow rapidly before
decaying, in many cases requiring many millions of terms before we begin to see
convergence.

\section{Rotational Relaxation Time}
\label{app:RotationalRelaxationTime}

As discussed in the main text, we find that the slowest relaxation time of an
active Rouse chain is
\[
    \tau_R = \frac{N^2}{\pi^2 + \alpha^2}.
\]
However, this relaxation is not necessarily tied to the dominant term in the sum
for $\ev{\vb{L}(t + \tau) \cdot \vb{L}(t)}$.

We measure the rotational relaxation time in the simulations by computing
$C_{\vb{L}}(\tau) = \ev{\vb{L}(t + \tau) \cdot \vb{L}(t)}$ and finding the time
$\tau^*$ at which $C_{\vb{L}}(\tau) = 1 / e$. Assuming $C_{\vb{L}}(\tau) =
\exp(-\tau / \tau_R^\text{eff})$, we find that $\tau^* = \tau_R^\text{eff}$,
where $\tau_R^\text{eff}$ is the effective rotational relaxation time as
measured from simulations. Comparisons of $\tau_R$ and $\tau_R^\text{eff}$ are
shown in Fig.~\ref{fig:RotationalRelaxationTime}.

\end{document}